\documentclass{emulateapj}
\usepackage{natbib}
\usepackage{ifpdf}
\slugcomment{Accepted for publication in ApJ}

\begin{document}
\newcommand\etal{et al. }
\newcommand\msun{{\,M_\odot}}
\newcommand\rsun{{\rm \,R_\odot}}
\newcommand\zsun{{\rm \,Z_\odot}}
\newcommand{\unit}[1]{\ensuremath{\, \mathrm{#1}}}

\shorttitle{THE FIRST GALAXIES}
\shortauthors{JEON ET AL.}

\title{THE FIRST GALAXIES: ASSEMBLY WITH BLACK HOLE FEEDBACK}

\author{
Myoungwon Jeon\altaffilmark{1},
Andreas H. Pawlik\altaffilmark{1},
Thomas H. Greif\altaffilmark{2},
Simon C. O. Glover\altaffilmark{3},
Volker Bromm\altaffilmark{1,2},
Milo\v s Milosavljevi\'{c}\altaffilmark{1},
Ralf S. Klessen\altaffilmark{3}
}
\altaffiltext{1}{Department of Astronomy and Texas Cosmology Center, University of Texas, Austin, TX 78712, USA; myjeon@astro.as.utexas.edu}
\altaffiltext{2}{Max-Planck-Institut f\"ur Astrophysik, Karl-Schwarzschild-Strasse 1, 85740 Garching bei M\"unchen, Germany}
\altaffiltext{3}{Zentrum f\"ur Astronomie der Universit\"at Heidelberg, Institut f\"ur Theoretische Astrophysik,
Albert-Ueberle-Strasse 2, 69120 Heidelberg, Germany}

\begin{abstract}
We study how the first galaxies were assembled under feedback from the
accretion onto a central black hole (BH) that is left behind by the
first generation of metal-free stars through self-consistent,
cosmological simulations. X-ray radiation from the accretion of gas onto BH
remnants of Population~III (Pop~III) stars, or from high-mass X-ray
binaries (HMXBs), again involving Pop~III stars, influences the mode 
of second generation star formation. We track the evolution of the
black hole accretion rate and the associated X-ray feedback starting
with the death of the Pop~III progenitor star inside a minihalo and
following the subsequent evolution of the black hole as the minihalo
grows to become an atomically cooling galaxy. We find that X-ray
photoionization heating from a stellar-mass BH is able to quench
further star formation in the host halo at all times before the halo enters 
the atomic cooling phase. X-ray radiation from a HMXB,
assuming a luminosity close to the Eddington value, exerts an even
stronger, and more diverse, feedback on star formation. It photoheats the gas inside 
the host halo, but also promotes the formation of molecular hydrogen and cooling of gas 
in the intergalactic medium and in nearby minihalos, leading to a
net increase in the number of stars formed at early times. Our simulations further
show that the radiative feedback from the first BHs may strongly
suppress early BH growth, thus constraining models for the
formation of supermassive BHs.
\end{abstract}


\section{Introduction}
\label{sec:intro}


Radiative feedback from the first generation of stars, the so-called
Population~III (Pop~III), is a crucial ingredient in determining how
the first protogalaxies assembled, and in setting the initial
conditions for subsequent, second-generation star formation 
\citep{Barkana2001, Bromm2004, Ciardi2005, Bromm2009}. It has been a longstanding goal to understand the formation
of the first galaxies subject to different accompanying feedback
effects from the first stars, such as ionizing feedback from
individual Pop~III stars \citep{Ricotti2002, Abel2007, Yoshida2007a}, 
chemical feedback produced by a supernova
(SN) explosion \citep{Wise2008, Greif2010, Wise2012}, and radiative
feedback from accreting black holes (BHs) \citep{Kuhlen2005,Alvarez2009, Park2011}.
The radiative feedback from accreting BHs on the formation of the first stars and the assembly of the 
first galaxies is the focus of the current  work.

The first stars are expected to form at a redshift $z\gtrsim 15$
inside dark matter minihalos with masses of $\sim 10^{6}\msun$ \citep{Haiman1996,
Tegmark1997,Yoshida2003}. Since hydrogen
molecules were the only effective low-temperature coolant in the
absence of metals, the primordial gas remains relatively warm, with
typical temperatures around 200 K. As a consequence of this increased
thermal pressure, the first stars are expected to be very massive
\citep{Bromm1999, Bromm2002, Nakamura2001, Abel2002, Omukai2003, Yoshida2006, OShea2007}. 
These Pop~III stars produce copious amounts of ionizing UV photons,
which can photoheat and evacuate the gas residing within the host
minihalos \citep{Whalen2004, Alvarez2006, JGB2007}. 

After their short lifetime of $\sim3$~Myr, Pop~III stars
end their lives in a SN explosion, or the direct collapse into a BH,
depending on the progenitor mass \citep{Heger2003}. If the
progenitor's mass lies in the range of $15-40\msun$, a conventional
core-collapse SN occurs, or, in the case of rapid rotation, a
hypernova. The resulting chemical abundance patterns are in good
agreement with those observed in extremely metal-poor stars in the
Galactic halo \citep{Beers2005,Joggerst2010}. 

An extreme fate would be a pair-instability supernova (PISN), predicted
for progenitor masses between $140-260\msun$ \citep{Barkat1967, Heger2002}. 
A consequence of the very large PISN yields
would be that even a single PISN could enrich $\sim10^7\msun$ of cold,
dense gas up to $10^{-4}-10^{-3}\zsun$ \citep{Karlsson2008, Greif2010}. 
Such high metallicities would be in excess of any predicted
critical value, beyond which a transition of star formation mode from
massive Pop~III stars to normal Population~II (Pop~II) would occur 
\citep{Omukai2000,Bromm2001b, Schneider2002}.

For progenitor masses within the range of $40\msun \lesssim M \lesssim
140\msun$ or $M \gtrsim 260\msun$, the Pop~III star will become a
massive BH via direct gravitational collapse. The ensuing merging of
minihalos, provided they contain cold gas, and the smooth accretion of
gas from the intergalactic medium (IGM) will feed the central BH. A fraction of the accreted
mass-energy will then be released as radiative energy, resulting in a
miniquasar \citep{Madau2001, Ricotti2004, Madau2004, Kuhlen2005, Wheeler2011}, which 
will ionize the surrounding neutral medium and heat the pre-galactic gas. 

Recent simulations have shown that, even in the absence of metal-line
and dust cooling, the primordial gas is able to fragment, leading to a
broad initial mass function (IMF) including also primordial stars of relatively low mass \citep{Turk2009, 
Stacy2010, Clark2011a, Clark2011b, Greif2011, Smith2011, Prieto2011, Stacy2012}, which maybe organized 
in binaries or stellar systems of still higher multiplicity. Such a scenario raises the possibility of the presence of high-mass X-ray binaries
(HMXB) at $z \gtrsim 10$, i.e., binary systems composed of a black
hole accreting gas from the surface of its stellar companion \citep{Mirabel2011, Haiman2011}. 
HMXBs may provide a much stronger feedback on the formation of the first galaxies than isolated
BHs, which have X-ray luminosities that depend crucially on the 
densities and temperatures of the surrounding gas \citep{Kuhlen2005,Alvarez2009, Park2011}. 

We should note that the assumed presence of HMXBs at high redshifts 
sensitively depends on the Pop~III IMF and multiplicitly. Recent simulations
have provided intriguing hints for the Pop~III IMF such that the Pop~III stars were likely to 
form in binaries or multiples rather than in isolation, but any predictions are subject to uncertainties
due to the limitations of the sink particle method. A sink particle treatment, which has been commonly adopted in smoothed particle hydrodynamics (SPH) simulations to follow the subsequent evolution of a protostellar core, allows us to avoid the 
problem of prohibitively small numerical timesteps, the so-called `Courant myopia' \citep{Bate1995, Bromm2002, Jappsen2005}. The resulting protostellar IMF is affected by the detailed treatment for sink particle mergers and by the ability of such schemes to self-consistently model the interaction between sink particles and the surrounding gas \citep{Greif2011}. It is, therefore, still an open question how the early protostellar mass function will be mapped into the 
final IMF of Pop III stars.

In this paper, we
investigate the feedback effects from such accreting stellar-mass
BHs by focusing on the question: ``How does a stellar black hole, a
remnant of a Pop~III star, influence the subsequent star formation and
in turn the assembly process of the first galaxies?" Our work is complementary to 
\citet{Greif2010} because our cosmological simulations start from the same initial conditions. 
The difference is that we include the feedback from accreting BHs while in their work the feedback from a PISN 
was taken into account. We can thus investigate how the assembly process of the
first galaxies differs in response to different assumptions regarding
the fate of the first stars and the accompanying feedback. In addition, we can constrain the
resulting gas properties inside the centers of such primordial
galaxies, thus providing us with the initial conditions for
second-generation star formation.

It is challenging to unambiguously define the first galaxies \citep[e.g.,][]{Bromm2011}. 
A plausible approach is to posit that the first galaxies should be massive enough to exhibit a sufficiently deep
potential well to retain gas that is photoheated by stars or
BHs. Another criterion is that a galaxy should sustain a
self-regulated star formation mode. Recent cosmological simulations
suggest that the first galaxies resided in so-called atomic cooling
halos \citep{Oh2002} with typical masses $\sim 10^8\msun$ at
$z\sim10$ \citep{Wise2007, Wise2008, Greif2008, Greif2010}. Whether
the presence of an X-ray radiation field such as produced by accreting
BHs will result in positive or negative feedback on the
formation of these galaxies remains an open key question \citep{Oh2001, Ricotti2001,
Glover2003, Machacek2003, Ricotti2004, Kuhlen2005, Zaroubi2007, Ripamonti2008,Thomas2008}. 
A positive feedback would arise if $\rm H_2$
molecule formation were promoted due to the elevated fraction of free electrons
from X-ray photoionization, facilitating the cooling and collapse of
gas and thus star formation inside minihalos. Alternatively, X-ray
heating may have suppressed star formation by increasing the entropy
floor in the pre-galactic medium, implying a negative feedback.

\begin{table}[t]
\caption{Simulation Parameters}
\centering
\begin{tabular}{c c c}
\hline\hline
Simulation & BH feedback & BH multiplicity \\
\hline
BHN & No & - \\
BHS & Yes & BH single (isolated) \\
BHB & Yes & BH binary (HMXB) \\
\hline
\\
\end{tabular}
\label{table:simul}
\end{table}

Miniquasar feedback may also be expected to have important effects on BH growth. One
outstanding question is how the supermassive black holes (SMBHs),
observed by the \emph{Sloan Digital Sky Survey} (SDSS) at $z\sim6$
with masses of $\sim10^9\msun$ \citep{Fan2006}, were able to grow within
such a short period of time after the Big Bang \citep{Haiman2001}. 
To answer this question, some possible scenarios have been
suggested, depending on the initial seed mass \citep{Bromm2011}. 
One of the main ideas is that the SMBH is born as a
stellar-mass BH, the remnant of a Pop~III star with a mass close to
that of its progenitor, and grows through the accretion of surrounding
gas and mergers with BHs that form in neighboring halos \citep{Madau2001, 
Li2007,Volonteri2006,Tanaka2009}. However, the photoheating and photoionization 
from a massive Pop~III star is strong enough to significantly suppress further supply
of cold gas onto the BH, creating a serious early bottleneck to growth 
\citep{JB2007,Pelupessy2007, Alvarez2009, Park2011}.

A competing scenario suggests the direct collapse of primordial gas
into an atomic cooling halo in the absence of low-temperature coolants
such as $\rm H_2$, or metals and dust. Due to the lack of low-temperature
coolants, the fragmentation of the gas will be avoided, and the
ensuing isothermal collapse may result in the direct formation of a
BH without Pop~III stellar progenitor. As a result, the mass of the seed BH is on the order of
$10^4-10^5\msun$, allowing subsequent growth from a more massive
foundation \citep{Bromm2003, Koushiappas2004, Wise2008, Regan2009, Schleicher2010, 
Shang2010,Johnson2011,Latif2011}. In this scenario,
negative feedback due to photoheating and radiation pressure may still
oppose growth \citep{Johnson2011}. A second goal of this study is to
provide an improved understanding of whether a stellar BH can grow to
become a SMBH in the presence of stellar and BH feedback.

Here we present the results of cosmological simulations, starting from
the point where Pop~III stars are first expected to form, up to the
virialization of a primordial galaxy, taking into account both the
ionizing radiation emitted by Pop~III stars and by miniquasars. We
assume that all Pop~III stars end their lives as massive BHs without SN, allowing
us to investigate solely the feedback from miniquasars and to compare
our results with the case where Pop~III stars die as PISNe. The outline
of this paper is as follows. In Section~\ref{NM}, we describe the numerical
methodology for the initial setup, the stellar radiative feedback, and
the feedback from miniquasars. In Section~\ref{FGA}, we show how the
primordial galaxy is assembled with and without BH feedback. Then, in
Section~\ref{PGP} we discuss the properties of the gas as it is falling into
the center of the galaxy. We present our conclusions in Section~\ref{SC}.

In the following, all lengths are expressed in physical (i.e., not comoving) units unless
explicitly stated otherwise.

\begin{figure}[t]
\epsscale{1.2}
\plotone{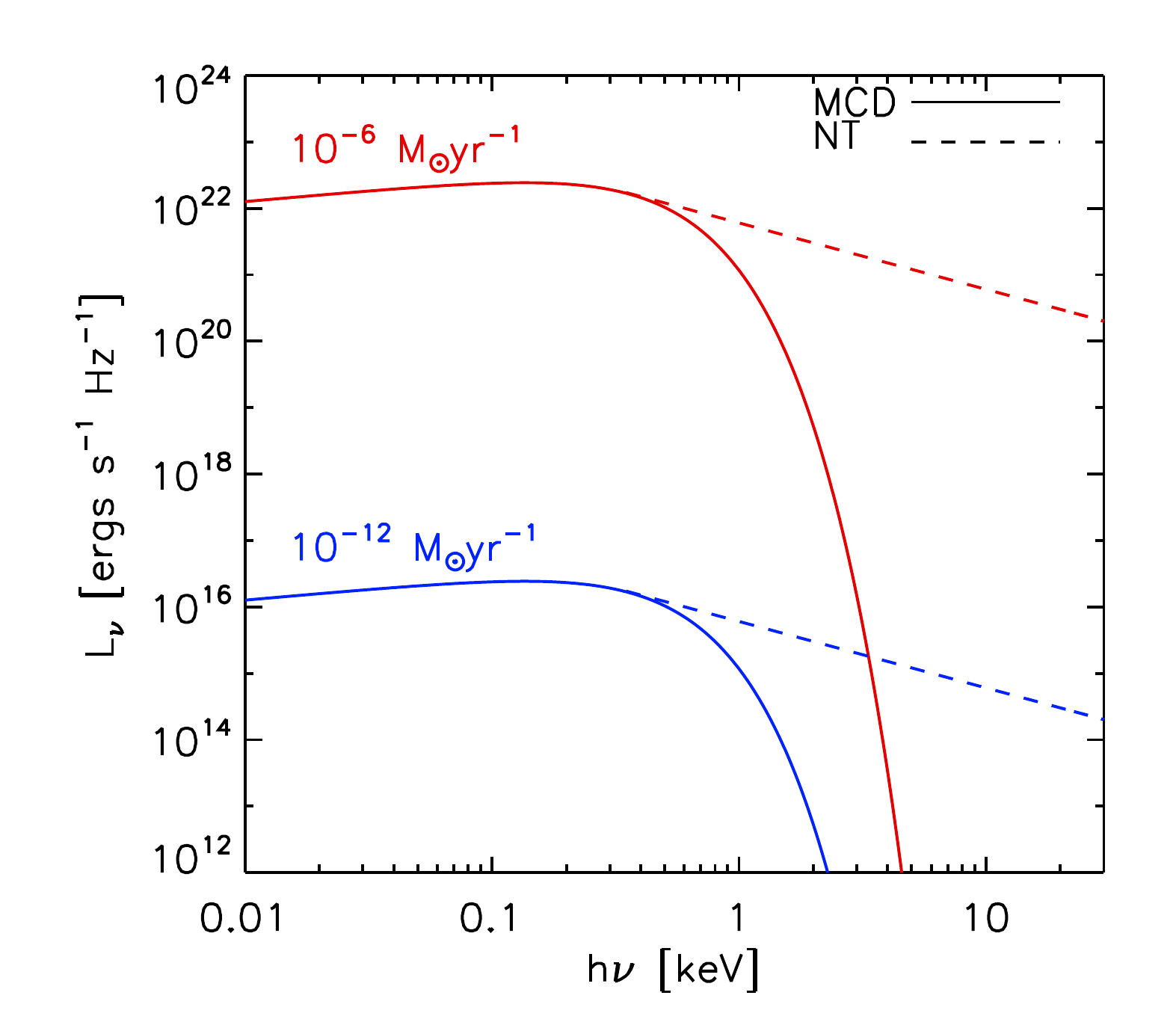}
\caption
{The emission spectra produced by gas accreting onto a 100$\msun$ BH
with two different accretion rates, $\dot{M}=10^{-6}\msun\rm yr^{-1}$
and $\dot{M}=10^{-12}\msun\rm yr^{-1}$. ``MCD'' (solid) shows the
continuum spectra from the multi-color disks and ``NT'' (dashed)
denotes a non-thermal component with a power law of $\beta = 1$. The
combined spectra are normalized to the total luminosity, $L = \epsilon
\dot{M} c^2$, where $\epsilon=0.1$ is the radiative efficiency,
assuming that each component contributes 50$\%$ to the total. Note that in
our simulations we ignore the multicolor disk component and instead assume
that the radiation emitted by the accreting BH is characterized by a pure power law spectrum.}
\label{spectra}
\end{figure}

\section{NUMERICAL METHODOLOGY}
\label{NM}

In this section, we describe the initial setup of the simulations and our methodology for calculating
the accretion rate onto BHs, as well as our treatment of radiative feedback from the Pop~III progenitor stars, an isolated BH remnant, and from a HMXB.

\subsection{Initial Conditions}
To survey the relevant parameter space, we have carried out three
cosmological simulations. As a reference, simulation ``BHN" includes
stellar radiative feedback from Pop~III stars, whereas the subsequent
feedback due to BH accretion is not taken into account.  The
simulation ``BHS" includes both the feedback from Pop~III stars and a
single isolated BH. Finally, in simulation ``BHB" we assume that the
Pop~III BH remnant has a stellar binary companion, giving rise to a
HMXB. We provide a summary of the simulations in
Table~\ref{table:simul}.

For the simulations presented here, we use a parallel version of the combined hydrodynamics and tree
N-body code GADGET2 \citep{Springel2005}. The code evaluates hydrodynamical forces
using the SPH technique.
Our simulations are initialized using a snapshot from the earlier simulation of \citet{Greif2010}.
We start running the simulations at $z\sim30$, corresponding to the point just before the first Pop~III
star in the computational box is formed, and terminate them at $z\sim10$, when the assembly of the first galaxy is expected to be complete.
The original simulation was initialized at $z = 100$ in a periodic box of linear size of 1 Mpc (comoving), using
$\Lambda$CDM cosmological parameters with matter density $\Omega_m=1-\Omega_{\Lambda}=0.3$,
baryon density $\Omega_b=0.04$, present-day Hubble expansion rate $H_0 = 70\unit{km s^{-1} Mpc^{-1}}$,
spectral index $n_{\rm s}=1.0$, and normalization $\sigma_8=0.9$,
consistent with the WMAP5 measurements \citep{Komatsu2009}. Given that we use a higher value compared to the current measurement of $\sigma_8=0.8$ \citep{Komatsu2011}, structure formation may be accelerated in our simulations. However, 
we compensate to some extent for the lack of large-scale power in our simulations due to the limited box size, by taking a value larger than the cosmological one. We also stress that in this paper we focus on individual halo properties which are not sensitive to the variation in the $\sigma_8$ parameter.    

Using a standard zoom-in technique \citep[see][]{Greif2010}, a preliminary
run with 64$^{3}$ particles was hierarchically refined to generate initial conditions
with high$-$mass resolution inside the region destined to collapse into the first galaxy. Employing three
consecutive levels of refinement, the mass of DM and gas particles in the highest resolution region
is $m_{\rm DM}\sim$33$\msun$ and $m_{\rm sph}\sim$5$\msun$, respectively. The corresponding baryonic  mass resolution is
$M_{\rm res}\sim 1.5 N_{\rm neigh} m_{\rm sph}\sim 400\msun$
, where $N_{\rm neigh}\sim 50$ is the number of neighboring particles within the SPH smoothing kernel \citep{Bate1997}. The Jeans mass in primordial gas, where molecular hydrogen cooling
imprints a characteristic density of
$n_{\rm H}=10^4 \unit{cm^{-3}}$ and a temperature of $200$ K \citep{Bromm2002}, is thus
marginally resolved.

\subsection{Chemistry, Heating, and Cooling}
We use the same chemistry and cooling network as in \citet{Greif2010}, where all relevant cooling mechanisms, such as H and He collisional ionization, excitation and recombination
cooling, bremsstrahlung, inverse Compton cooling, and collisional excitation cooling via $\unit{H_2}$
and HD, are taken into account. For $\unit{H_2}$ cooling, collisions with protons and electrons are explicitly included, in
addition to the usually dominant neutral hydrogen atoms.
The code self-consistently solves the rate equations for the abundances of $\rm H, H^+, H^-, H_2, H_2^+,
He, He^+, He^{++}$, and $\unit{e^-}$, as well as the three deuterium species D, $\rm D^+$, and HD.
We can thus accommodate the non-equilibrium chemical evolution which is ubiquitous in early
universe structure formation.

\subsection{Sink Particle Method}
\label{s23}

Employing the sink particle algorithm of \citet{JB2007}, we
convert an SPH particle into a collisionless sink particle if its
hydrogen number density exceeds a threshold value of $n_{\rm
max}=10^4$\,cm$^{-3}$. When a sink particle forms, gas particles
within an accretion radius, $r_{\rm acc}$, are immediately accreted
onto the sink. The position and velocity of a sink particle are
estimated every timestep based on a new mass-weighted position and
velocity of the accreted gas particles within $r_{\rm acc}=L_{\rm
res}$. Here, the resolution length of the simulation is defined as:
\begin{equation}
	L_{\rm res} = 0.5 \left( \frac{M_{\rm res}}{\rho_{\rm max}}
	\right)^{1/3} \simeq 1 \unit{pc},
\end{equation}
where $\rho_{\rm max} \simeq n_{\rm max} m_{\rm H}$.

A sink particle can grow in mass through the further accretion of
surrounding gas. The criterion for subsequent accretion is that
a neighboring SPH particle approaches the sink to within $r_{\rm acc}$, which
we hold constant throughout the simulation. Note that our simple
prescription for sink growth is sufficient for the purpose of providing a
marker for the position of a Pop~III star and its BH remnant. When determining the mass growth
of the BH which is responsible for the X-ray feedback, we use a more
sophisticated methodology to determine its accretion rate (see Section~\ref{s25}).

\begin{figure}[t]
\epsscale{0.95}
\plotone{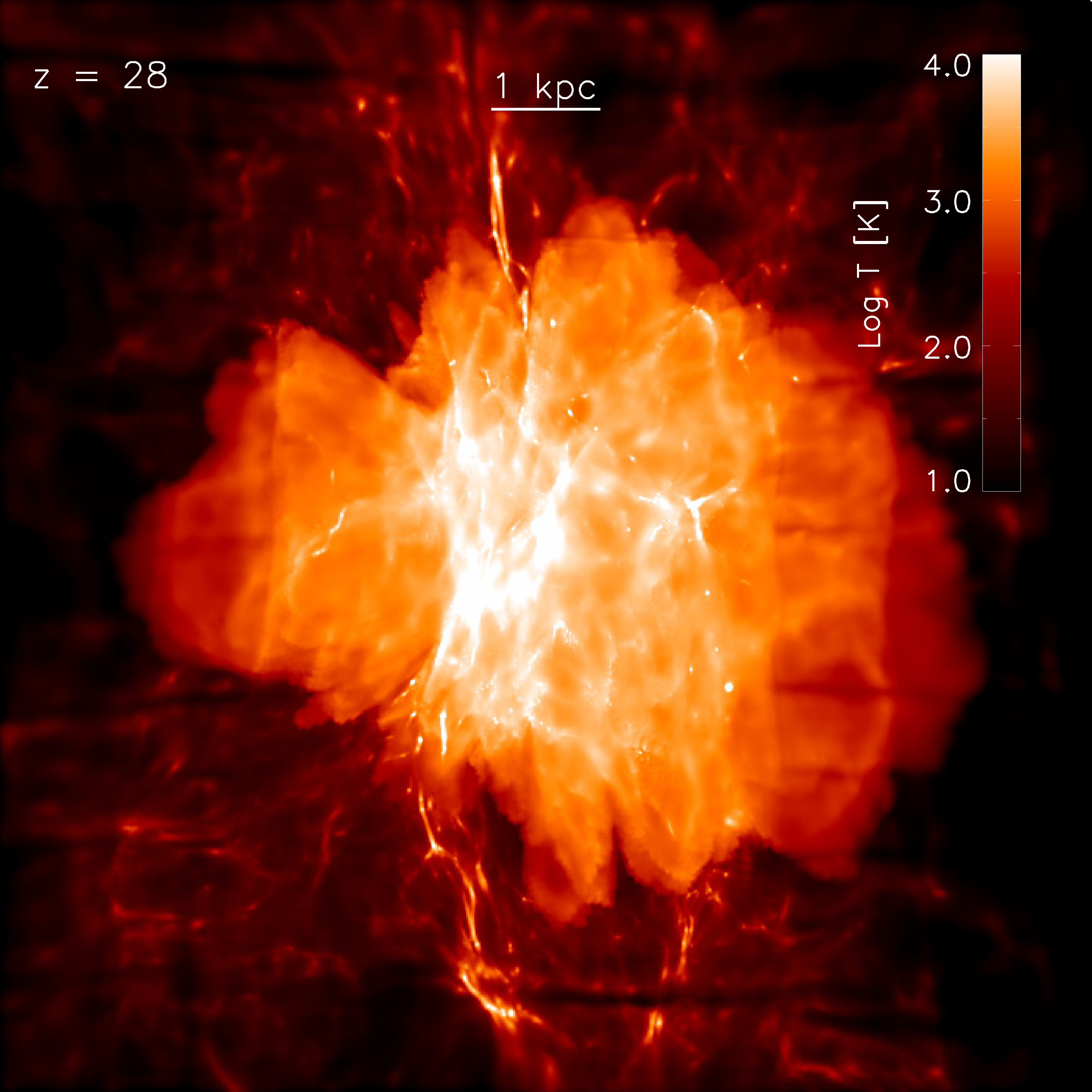}
\caption
{Radiative feedback from the first star. We show the projected temperature distribution, employing a density-squared weighting scheme, of the \ion{H}{2} region around a $100 M_{\odot}$ Pop~III star at $z=28$, briefly after its death. The size\protect\footnotemark[1] of the final \ion{H}{2} region, $\sim5$ kpc, greatly exceeds the virial radius of the host halo, $\sim$100 pc.}
\label{HII}
\end{figure}

\footnotetext[1]{A numerical error in the code for the ionization front tracking caused
the simulations presented in this work to overestimate the
radii of the Pop III H~II regions by a factor $\sim$ 2, implying final
H~II region radii of $\sim$ 5 kpc instead of $\sim$ 2-3 kpc
\citep[compare with, e.g.,][their Figure~3]{Greif2009}. However, this overestimate of the 
radii does not affect the conclusions of the present work. 
This is because we are discussing the differences between simulations with 
and without black hole feedback, which was correctly implemented, and because the radii of
the H~II regions were overestimated in a systematic manner by the same
factor in all these simulations. In any case, the simulations are consistent with the case 
including feedback from more massive Pop~III stars with larger \ion{H}{2} regions.}

\subsection{Stellar Radiative Feedback}
\label{s24}
After a Pop~III star has formed inside its host minihalo, we turn on the radiation field emitted by this star, treated as a point source.
We propagate the ionization front around it to build up a primordial \ion{H}{2} region,
using a well-tested ray-tracing algorithm \citep{Greif2009}. This scheme
solves the ionization front equation in a spherical grid by tracking $10^5$ rays with 500
logarithmically spaced radial bins around Pop~III stars. The hydrodynamical effect is taken into account by
self-consistently coupling the ray-tracing module to the chemical and thermal evolution of the gas.

We assume that a $100 \msun$ Pop~III star emits blackbody radiation
with an effective temperature
$T_{\unit{eff}}=10^{4.9}$\,K, and luminosity $L_{\ast}=10^{6.1} L_{\odot}$ \citep{Bromm2001a, Schaerer2002}.
The corresponding production rates for ionizing photons are: $\dot{N}_{\rm ion, HI/HeI}=9.1 \times 10^{49} \unit{s^{-1}}$
and $\dot{N}_{\rm ion, HeII}=4.1 \times 10^{48} \unit{s^{-1}}$. The growth of the ionization front continues
until it reaches its maximum size at the end of the Pop~III star's life after $t_{\ast}=$ 2.7 Myr.
However, not all Pop~III stars are endowed with an \ion{H}{2} region because of the computational expense incurred by the ray-tracing.
We trace the photons only from those stars formed within the Lagrangian volume destined to become the first galaxy at $z\sim 10$, roughly corresponding
to a 10~kpc radius from the center of this region. 

Finally, we include the transfer of the $\rm H_2$-dissociating Lyman-Werner (LW) photons
in the range of 11.2$-$13.6 eV, isotropically emitted by a Pop~III star as a $1/r^2$ field without attenuation of the flux with radius, which propagate far beyond the H~II region. The photo-dissociating rate is $k_{\rm H_2} = 1.1 \times 10^8 F_{\rm LW}$, where $F_{\rm LW}$ is the radiation flux integrated over the LW bands \citep{Abel1997}. We compute $F_{\rm LW}$ in the optically thin limit and do not take $\rm H_2$ self-shielding into account \citep[see][for more details]{Greif2009}. The neglect of self-shielding is justified due to the low H$_2$ column densities encountered in our simulations, of order $\sim10^{13} \unit{cm^{-2}}$ inside the BH host galaxy \citep{Draine1996, Wolcott2011}.


\subsection{Black Hole Feedback}
\label{s25}

In an important difference from \citet{Greif2010}, we here assume
that all Pop~III stars die via direct collapse into BHs, without any preceding SN explosion. For simplicity, however, we allow only one BH to produce radiative feedback due to the accretion onto it, taken as the first BH formed during the simulation. The heat input due to this BH feedback into the pre-galactic region is active for $\sim$ 350 Myr, while the radiation from individual Pop~III stars is turned on for only 2.7 Myr.
We next describe our treatment for estimating the accretion rate onto a BH and the model for the radiation
emitted by an isolated BH and a HMXB.


\begin{figure*}[ht]
\epsscale{1.2}
\plotone{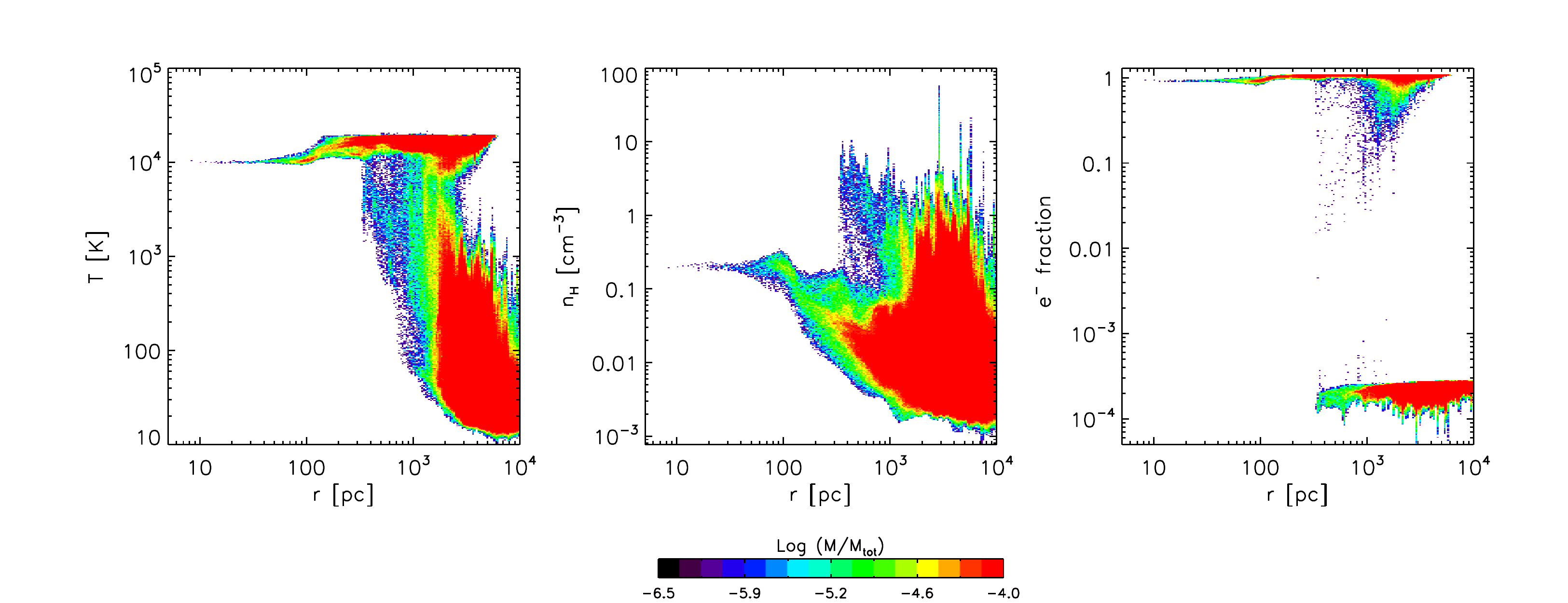}
\caption
{Gas properties within the \ion{H}{2} region depicted in Fig.~\ref{HII}. We show the situation
$\sim$1 Myr after the radiation from the Pop~III star was switched off. From left to right, the panels show gas temperature, hydrogen number density, and electron fraction as a function of distance from the central star. The color coding indicates the fraction of mass per pixel in a given phase. The total gas mass, $M_{\rm tot}$, corresponds to the central 10 kpc.}
\label{HII_gas}
\end{figure*}

\subsubsection{Accretion Rate}
We assume that the BH is embedded in a pressure-supported primordial gas cloud, steadily accreting from it. The corresponding accretion rate is given by the Bondi \& Hoyle \citeyearpar{Bondi1944} model, where a homogeneous medium which is
at rest at infinity accretes onto a point mass. The Bondi-Hoyle rate can be written as:
\begin{equation}
\label{bondiM}
    \dot{M}_{\rm BH} = \frac{4\pi(GM_{\rm BH})^2 \rho_{\rm gas}}{(c_s^2+v_{\rm BH}^2)^{3/2}},
\end{equation}
where the gas sound speed, $c_s$, and gas density, $\rho_{\rm gas}$, are determined by averaging over the $N_{\rm neigh}\simeq 50$ SPH particles closest
to the BH. The relative speed of the BH with respect to
the surrounding gas, $v_{\rm BH}$, is negligibly small throughout the simulation. Varying $N_{\rm neigh}$ by an order of magnitude does not affect the accretion rates when they are estimated with Equation~(\ref{bondiM}).

By assuming that 10\% of the rest-mass energy of the accreted matter
is released as radiation, we normalize the total luminosity according
to $L = \epsilon \dot{M}_{\rm BH} c^2$, where $\epsilon=0.1$ is the
radiative efficiency and $c$ the speed of light.  We estimate that
the accretion rate $\dot{M}_{\rm BH}$ lies between $10^{-12}\msun \rm yr^{-1}$ and $10^{-6}\msun \rm yr^{-1}$. The former corresponds to the initial situation, where the BH
is located at the center of an \ion{H}{2} region with a temperature of $\sim 10^4$\,K and hydrogen number density of $\sim \rm
1$\, cm$^{-3}$. The upper limit is derived from comparison
simulations where stellar and BH radiative feedback is neglected.

It is important to
note that the accretion rate used in this work is an upper limit, for three reasons. 
In reality, when also the feedback from 
radiation pressure is taken into account, the true rates
are likely much smaller than the nominal Bondi-Hoyle value 
\citep{Milos2009a, Milos2009b, Park2012}. We
estimate, assuming a typical accretion rate of $10^{-8}\msun \rm yr^{-1}$, that the acceleration due to radiation pressure from Thomson
scattering and photoionization in the vicinity of the BH, at $r =
1$\,pc, is 30 times larger than the gravitational acceleration,
$a_{\rm grav}=GM_{\rm BH}/r^2$ \citep[e.g., Eq.~10 in][]{Johnson2011}. This indicates that radiation
pressure, mainly due to photoionization, is important near the BH,
acting to limit growth. The other reason is that we have only
considered the case of radiatively efficient accretion. If the cooling
time scale for the liberation of viscously generated energy is longer
than the accretion time scale, most of the energy is advected inward
with the accreting gas instead of being radiated away \citep{Narayan1994, 
Narayan1995, Blandford1999, Blandford2004}. A fraction of the
energy carried by the outflow could ultimately be converted to
radiation, but the resulting spectrum would certainly be very
different than that of the thin disk assumed here. A third reason is that 
the vorticity of the turbulent gas has not been properly resolved here. Studying its effect on Bondi accretion, \citet{Krumholz2005} found that even a small amount of vorticity is able to significantly reduce the accretion onto the BH. This suppression could be important in an atomic cooling halo, where the radial, cold flow along the filaments is converted into turbulent motion, thus generating vorticity in the central region of the galaxy.

In addition to the Bondi-Hoyle prescription, there is another way to
estimate BH growth, based on $\dot{M}_{\rm sink}$ calculated with the
method in Section~\ref{s23}, where the BH mass grows by accreting all gas
particles within $r_{\rm acc}$. The resulting rate is much higher than
the Bondi-Hoyle value, roughly by a factor of 10. We adopt
$\dot{M}_{\rm BH}$ rather than $\dot{M}_{\rm sink}$ as an estimator
for BH growth, given that the sink accretion radius is about one order of
magnitude larger than the Bondi-Hoyle radius, $r_{\rm B} = (\mu m_{\rm
H} G M_{\rm BH})/k_{\rm B} T$, where $\mu$ is
the mean molecular weight and $k_{\rm B}$ is the Boltzmann constant, or, 
$r_{\rm B} \sim 0.3 \unit{pc}\left(M_{\rm BH}/10^2\msun\right)\left(T/10^2\unit{K}\right)^{-1}$.
Moreover, only a fraction of the infalling gas will likely reach the vicinity
of the BH. Furthermore, we can thus track the change in BH mass in a
smooth manner, and do not have to contend with discreteness effects
owing to the limited mass resolution afforded by SPH.


\subsubsection{Non-Thermal Radiation}
 In the local universe, the emergent spectra from accreting BHs are
 typically modelled as a combination of a power-law component,
 $F_{\nu}\propto \nu^{\beta}$, describing non-thermal (NT) synchrotron
 radiation, and a multi-color disk (MCD), resulting in a soft, thermal
 continuum \citep[e.g.,][]{Mitsuda1984}. In Figure~\ref{spectra}, we show typical
 spectra to illustrate this model where a soft thermal component is
 fitted by a MCD blackbody with $kT_{\rm in}\sim$ 150 eV \citep{Miller2003}, 
normalized assuming two representative accretion
 rates. Here, $T_{\rm in}$ is the temperature of the inner disk
 region which is related to the BH mass according to $T_{\rm
 in}\propto M_{\rm BH}^{-1/4}$. It is currently not known whether
 accreting BHs at high redshift would behave in a similar
 manner. 

 Conservatively, we here assume that there is no significant
 difference in the BH emission physics over cosmic time, and use the
 same spectra. However, we ignore the MCD component of the emitted 
 radiation in our simulations. Such a thermal
 contribution would determine the radiative feedback in the immediate
 ``near zone'' from the central BH, but is unimportant on larger
 scales, where the NT component dominates. We estimate the size of the
 near zone with a standard Str\"omgren analysis. Assuming that the
 ionizing photons originate from the MCD component, and employing a
 typical accretion luminosity of $\dot{M}_{\rm
 BH}=10^{-7}\msun$\,yr$^{-1}$ and a typical density of $n_{\rm H} =
 10$\, cm$^{-3}$, we compute an equilibrium ionization front (I-front) radius of
 $\lesssim 5$\,pc, which is comparable to the spatial resolution of the simulations.  

 We follow \citet{Kuhlen2005} to model the propagation
 of high-energy photons where the isotropic radiation field, $\propto
 1/r^2$, is a function of distance from the BH only. The coefficient for 
the ionization rate, $k_{\rm ion}$, can be written as

\begin{equation}
\label{ion2}
        k_{\rm ion} = \int_{\nu_{\rm NT}}^{\nu_{\rm max}} \frac{F_{\nu} \sigma_{\nu}}{h\nu} d\nu \\
\end{equation}

where $\nu_{\rm NT} = 0.2$\,keV$/h$ is the frequency where the non-thermal component starts to dominate and $\nu_{\rm max} = 10$\,keV$/h$ is the upper cutoff frequency. Our choice of energy range, $0.2-10$\,keV, is motivated by the study of \citet{Miller2003}, where the spectra of select ULXs associated with $100 \msun$ black holes were fitted with a power law in this same range.
For the hydrogen and helium photo-ionization cross sections, $\sigma_{\nu}$, we use the standard expressions \citep[e.g.,][]{Barkana2001, Osterbrock2006}.

\begin{figure}[t]
\epsscale{1.1}
\plotone{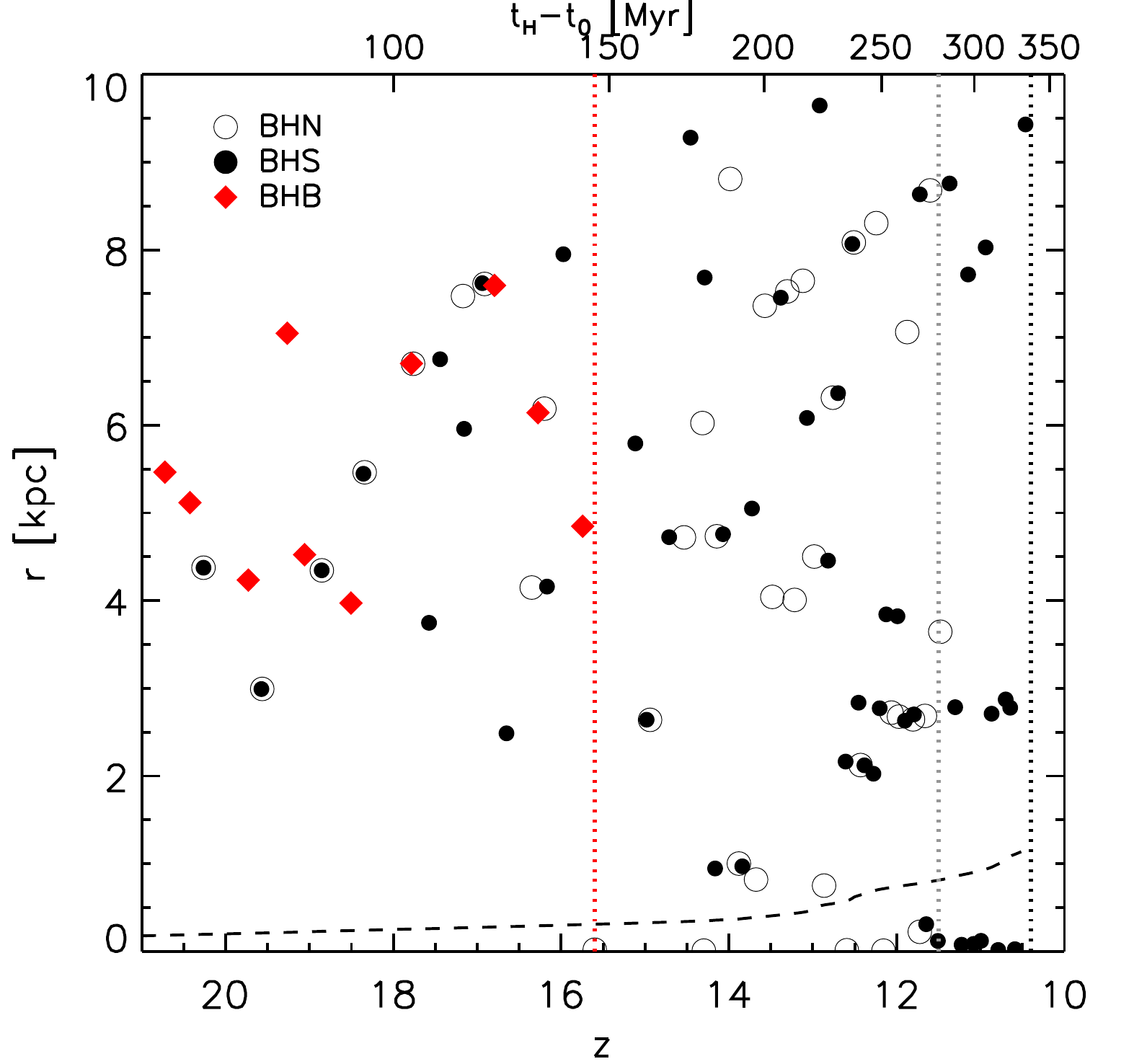}
\caption {Distances between newly formed Pop~III stars that are accompanied by \ion{H}{2} regions and the BH as a function of redshift from our three simulations. Note that the BHN, BHS and BHB simulations end at different redshifts, corresponding to $z\approx 11.6$, $z\approx 10.4$, and $z\approx 15.8$, respectively (indicated by the vertical dotted lines). A total of $\sim 50$ Pop~III stars form within 10 kpc around the central BH in the range of redshifts $z=20-10$ in simulations BHN and BHS. In the BHB run, the positive feedback is clearly evident far away from the source. This is due to the gas collapse into distant minihalos, facilitated via $\rm H_2$ cooling promoted by the strong X-ray emission from the HMXB; locally, on the other hand, star formation is suppressed due to the strong negative feedback from the binary source. We also show the virial radius of the DM halo hosting the active BH or the HMXB({\it dashed line}). The X-ray photoionization heating from a stellar-mass BH is also able to quench further star formation in the host halo at all times before the halo enters the atomic cooling phase. The top axis shows the time elapsed since the formation of the black hole at time $t_0$.}
\label{sf}
\end{figure} 

\begin{figure}[t]
\epsscale{1.2}
\plotone{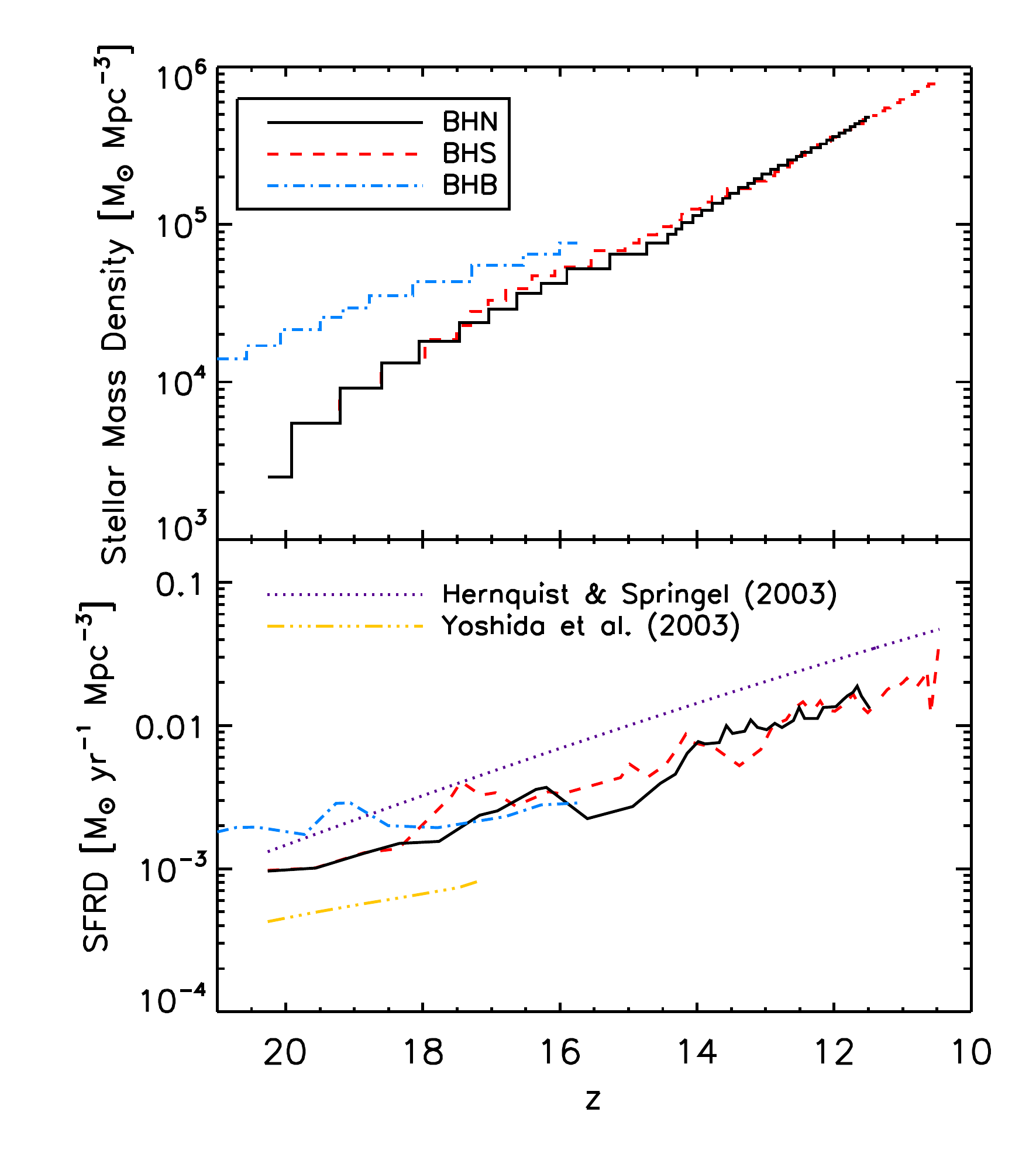}
\caption{\emph{Top panel}: Comoving stellar mass density  vs. redshift. We compare the results from our three simulations, the BHN comparison case (\emph{solid line}), the BHS feedback run (\emph{dashed line}), and the BHB run 
(\emph{dot-dashed line}). It is evident that the positive feedback from the BH binary, acting on cosmological scales,
 has boosted star formation at the highest redshifts. \emph{Bottom panel}: Comoving star formation rate density vs. redshift.
 The lines refer to the same simulations as above. For comparison, we overplot the analytic fitting formula, derived by \citet{Hernquist2003}, for higher-mass systems where atomic cooling is active (\emph{dotted line}), and that of \citet{Yoshida2003} for Pop~III star formation in minihalos via $\rm H_2$ cooling (\emph{dashed triple-dotted line}).}
\label{fig:sfr}
\end{figure}

To consider the secondary ionization effects from energetic electrons released through the absorption of an X-ray photon, we adopt
the fitting formulae given by \citet{Shull1985} \citep[see also][]{Valdes2008,Furlanetto2010}. They computed
the fractions of the initial electron energy going into secondary
ionizations of HI, secondary ionizations of HeI, and into heating the surrounding gas. Secondary ionizations of HeII are not 
important \citep{Shull1985}, and are thus not included in the overall energy budget. The effective ionization rates can
thus be written

\begin{equation}
\label{sec1}
        k_{\rm eff}[\rm HI] = k_{\rm ion}[\rm HI] + k_{\rm sec}[\rm HI] ,
\end{equation}
where
\begin{equation}
        k_{\rm sec}[\rm HI] =  f_{\rm H}\left(\Gamma_{\rm HI} + \frac{n_{\rm HeI}}{n_{\rm HI}}
 \Gamma_{\rm HeI}\right)\frac{1}{13.6 \unit{eV}} ,
\end{equation}
and
\begin{equation}
\label{sec2}
        k_{\rm eff}[\rm HeI] = k_{\rm ion}[\rm HeI] + k_{\rm sec}[\rm HeI] ,
\end{equation}
where
\begin{equation}
        k_{\rm sec}[\rm  HeI] =  f_{\rm He} \left(\Gamma_{\rm HeI} + \frac{n_{\rm HI}}{n_{\rm HeI}}
 \Gamma_{\rm HI}\right)\frac{1}{24.6 \unit{eV}}.
\end{equation}

Here $\Gamma$, the rate at which the excess energy is released due to the first photoionization of the gas by X-rays, is given by
\begin{equation}
\label{heat2}
        \Gamma = n_{\rm n} \int_{\nu_{\rm NT}}^{\nu_{\rm max}} F_{\nu} \sigma_{\nu}
        \left(1 - \frac{\nu_{\rm min}}{\nu}\right) d\nu,
\end{equation}

where $n_{\rm n}$ is the number density of the respective unionized species, $\nu_{\rm min}$ the ionization threshold frequency, $h\nu_{\rm min}$=13.6 eV, $h\nu_{\rm min}$=24.6 eV, and $h\nu_{\rm min}$=54.4 eV for hydrogen, neutral helium, and singly ionized helium, respectively.

The fractions going into secondary ionizations, $f_{\rm H}$ and $f_{\rm He}$, 
are a function of the hydrogen ionization fraction, $x_{\rm ion}=n_{\rm H+}/n_{\rm H}$, and can be expressed as
\begin{equation}
\label{f1}
 f_{\rm H} = 0.3908 (1-x_{\rm ion}^{0.4092})^{1.7592}
\end{equation}
\begin{equation}
\label{f2}
 f_{\rm He} = 0.0554 (1-x_{\rm ion}^{0.4614})^{1.6660}\mbox{\ ,}
\end{equation}
and the photo-heating rates are then $\mathcal{H}=\Gamma (1-f_{\rm H/He})$.

The flux from the non-thermal component is given by
\begin{equation}
\label{FNT}
 F_{\nu} = F_{\nu_0} \left(\frac{\nu}{\nu_{0}}\right)^{-\beta},
 \hspace{1cm} \beta = 1,
\end{equation}
in the range of 0.2$-$10 keV, implying a mean hydrogen-ionizing photon energy 
of 390 eV.  Here $F_{\nu_0}$, the flux at $\nu_0=\nu_{\rm NT}$, is determined
at every timestep, based on the normalization:
\begin{equation}
\label{FNT2}
 \epsilon \dot{M}_{\rm BH} c^2 = \int_{\nu_{\rm NT}}^{\nu_{\rm max}}
 L_{\nu} d\nu ,
\end{equation}
where, $L_{\nu}=4 \pi r^2 F_{\nu}$ is the specific luminosity. We
note that since extinction along the line of sight is not taken into
account, the estimated ionization and heating rates are likely upper limits. 
Recall that the accretion rate, $\dot{M}_{\rm BH}$, is
updated at every timestep, thus allowing our algorithm to reflect the
changing conditions in the vicinity of the BH.

The resulting photo-ionization and photo-heating rates are a function of distance from the BH, the instantaneous accretion rate, a function of the hydrogen ionization fraction, and can be conveniently written as

\begin{equation}
\label{ion1}
        k_{\rm ion} = \dot{K} \left(\frac{r}{\unit{pc}}\right)^{-2} \left(\frac{\dot{M}_{\rm BH}}{10^{-6}
     \msun/\rm yr}\right),
\end{equation}
where
\begin{equation}
	\dot{K} = [  1.96, 2.48, 0.49  ] \times 10^{-11} \hspace{0.5cm} \unit{s^{-1}} ,
\end{equation}
and
\begin{equation}
\label{heat1}
        \mathcal{H} = n_{\rm j} \hspace{0.1cm} \dot{H} \left(\frac{r}{\unit{pc}}\right)^{-2}
 \left(\frac{\dot{M}_{\rm BH}}{10^{-6}\msun/\rm yr}\right) \left(1-f_{\rm H/He}\right) \hspace{0.1cm},
\end{equation}
where
\begin{equation}
        \dot{H} = [  7.81, 9.43, 1.63  ] \times 10^{-21} \hspace{0.3cm} \unit{erg \hspace{0.05cm} s^{-1}}
\end{equation}
for neutral hydrogen, neutral helium, and singly-ionized helium, respectively. The corresponding number densities of each species $\rm j$ are given by $n_{\rm j}$.

\subsubsection{HMXB Emission}

Recent improved simulations of Pop~III star formation suggest that the
primordial gas could fragment into two or more distinct cores,
possibly resulting in a massive binary, or higher-multiple stellar
system \citep{Turk2009, Stacy2010, Stacy2012, Clark2011a, Clark2011b, Greif2011, 
Smith2011, Prieto2011}. This
picture provides another possible X-ray source connected to the first
stars, where the primary in a binary system leaves a BH behind at the
end of its short life \citep{Mirabel2011, Haiman2011}. The remnant could remain bound to the slightly
less massive secondary, where nuclear burning is still going on. The
material lost by the secondary in a wind or by Roche lobe overflow
will be dumped onto the companion BH, releasing copious X-ray
emission. The duration of such a HMXB is limited by the main-sequence
lifetime of the donor, and is of the order of $10^7$\,yr.

It has been suggested that there is a correlation between the number
of HMXBs and ultra-luminous X-ray sources (ULXs), which can be
explained as a subpopulation of the former, and low-metallicity 
environment \citep{Majid2004, Dray2006, Soria2007, Mapelli2009,
Linden2010}. We can thus expect that the number densities of HMXBs and
ULXs at high redshifts, when the universe was chemically pristine,
were higher than what is observed locally \citep{Mirabel2011,
Haiman2011}. Such local surveys show that HMXBs and ULXs have
luminosities of $L_X \sim 10^{38}$ erg $\rm s^{-1}$ and $L_X \gtrsim
10^{40} $erg $\rm s^{-1}$, respectively, which are close to the
Eddington value, or mildly super-Eddington
\citep{Grimm2003}. Consequently, we assume that the luminosity of a
HMXB at high redshifts is close to the Eddington limit, $L_{\rm
Edd}=4\pi G M_{\rm BH} m_{\rm H} c /\sigma_{\rm T}$, where $m_{\rm
H}$, $\sigma_{\rm T}$, and $c$ denote the proton mass, Thomson cross
section, and the speed of light, respectively. We further assume that
the spectrum exhibits the same power-law behavior as in
Equation~(\ref{FNT}). The X-ray
luminosity in the HMXB scenario is then determined from

\begin{equation}
\label{Edd}
        L_{\rm Edd} = 1.38 \times 10^{40} \unit{erg \hspace{0.1cm} s^{-1}} \left(\frac{{M}_{\rm BH}}{100\msun}\right) = \int_{\nu_{\rm NT}}^{\nu_{\rm max}} L_{\nu} d\nu,
\end{equation}
and we assume $M_{\rm BH}=100\msun$.

\section{First Galaxy Assembly}
\label{FGA}
In this section we present the results obtained from our
simulations. In Section~\ref{s31}, we briefly discuss the properties of the
\ion{H}{2} regions around massive Pop~III stars in order to check for
agreement between our work and previous studies. We proceed in
Section~\ref{s32} to derive the star formation rate in the Lagrangian volume
of the emerging primordial galaxy. We then discuss in Section~\ref{s33} how
this galaxy is assembled, contrasting the simulations of feedback from
a single isolated accreting black hole (BHS) and from a HMXB (BHB) 
described in Sec.~\ref{NM} with a simulation without BH feedback (BHN). 
We conclude with a description of the accompanying black hole growth.

Note that the BHN, BHS and BHB simulations end at different redshifts, 
corresponding to $z\approx 11.6$, $z\approx 10.4$, and $z\approx 15.8$, respectively. 
The reason that we terminated the BHN run earlier than the BHS case is mainly 
due to computational expense, but the galaxy at $z\approx 11.6$ already meets the criterion 
for first galaxies and the results give us enough information to compare with those of 
the BHS run. For the BHB case, we have performed the simulation for much 
longer than any realistic lifetime of HMXBs. For example, the main-sequence lifetime 
of a donor star of $30\msun$, close to the predicted typical Pop~III mass \citep{Clark2011b, Hosokawa2011, Stacy2012}, 
would be $\sim 20$\,Myr.  Thus, our work can be considered to represent the case where 
at least one HMXB exists continuously over 150 Myr, somewhere in the Lagrangian volume 
of the emerging protogalaxy.

\subsection{\ion{H}{2} Regions around the First Stars}
\label{s31}

The first Pop~III star to appear in our cosmological box forms at a
redshift $z=28$ inside a $5\times10^5\msun$ minihalo. Since the
Pop~III star is assumed to be very massive, with $M_{\ast}\simeq 100
\msun$, it produces copious amount of ionizing photons which carve out
an extended \ion{H}{2} region into the surrounding IGM. The gas within
the \ion{H}{2} region is photo-heated to temperatures, $T\sim 3\times
10^4$\,K, in excess of the virial temperature of the minihalo, thus
triggering a hydrodynamic outflow. In Figure~\ref{HII}, we show the projected
temperature approximately 1~Myr after the radiation from the central
star was switched off. The terminal size of the \ion{H}{2} region is
about $\sim$5 kpc\footnotemark[1], such that the I-front extends much farther out than
the virial radius of the minihalo, $r_{\rm vir}\sim$100 pc. Note that
the anisotropic morphology of the \ion{H}{2} region reflects the
inhomogeneous density distribution in the neighboring IGM. We describe
the physical properties of the photo-ionized gas in Figure~\ref{HII_gas}, again
$\sim 1$\,Myr after the Pop~III star died.  At this time, the gas
towards the center has already begun to cool and recombine, leading to
a relic \ion{H}{2} region.

\begin{figure}[t]
\epsscale{1.2}
\plotone{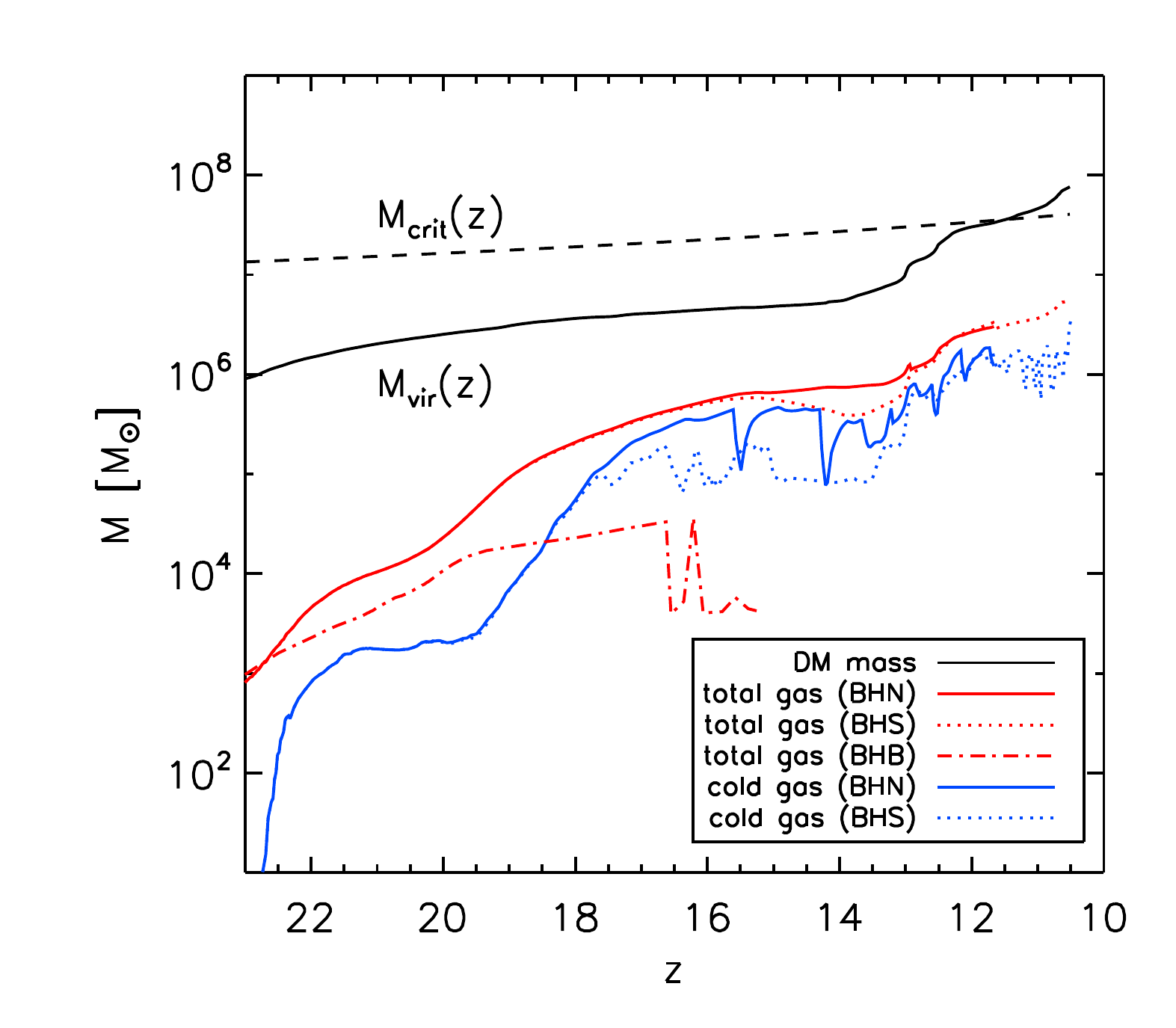}
\caption {Assembly of the first galaxy. Shown is the redshift evolution of DM virial mass (\emph{black solid line}) of the halo which will host the first galaxy at $z\sim10$, as well as of the total gas mass (\emph{red}) and the cold gas mass (\emph{cyan}) within the halo for simulations BHN (\emph{solid lines}), BHS (\emph{dotted lines}), and BHB (\emph{dot-dashed line}). The black dashed line represents the critical mass required for the onset of atomic cooling in the halo at a given redshift. There is no cold gas within the halo in simulation BHB owing to the strong heating from the HMXB.}
\label{fig:virial}
\end{figure}

\begin{figure*}[ht]
\epsscale{1.05}
\plotone{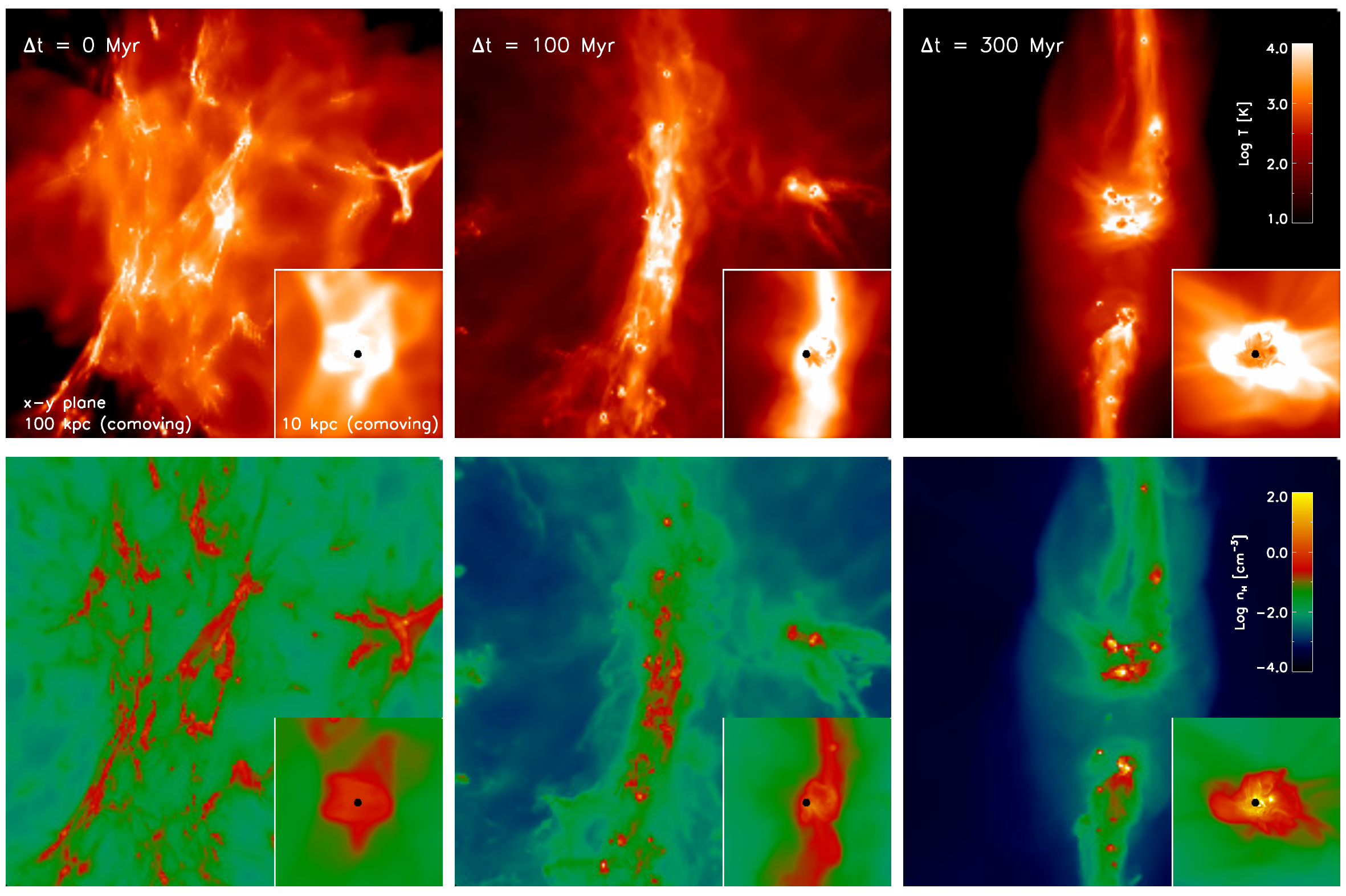}
\caption {Gas properties in the vicinity of the accreting BH in simulation BHS. Shown are the density-weighted temperature (\emph{top row}) and the projected hydrogen number density (\emph{bottom row}) of the gas within cubical slices of linear size 100 kpc (comoving). The small insets provide zooms into the central 10 kpc (comoving). The situation is depicted at three different times, $\Delta t$= 0, 100, and 300 Myr after the BH has been formed at the center of the minihalo, corresponding to $z=27.9$, 17.7, and 11.0, respectively. The location of the BH is marked by a black circle at the center of each plot.}
\label{multi}
\end{figure*}

The expansion of the I-front begins with a very short lived R-type
(supersonic) phase, followed by a D-type (subsonic) one, where the
I-front is trapped behind a hydrodynamical shock wave \citep[e.g.,][]{Whalen2004,Alvarez2006,JGB2007}. 
After $\sim10^5$ yr, the I-front is able to break out, now again supersonically
racing ahead of the shock into the low-density IGM. The resulting
density profile, together with the location of the shock at $\sim
$200\,pc, and a corresponding velocity of $v_{\rm sh}\sim 25-35$\, km
$\rm s^{-1}$, are in good agreement with the analytical champagne-flow
solution \citep{Shu2002}. We note that contrary to the minihalos of
$\sim10^6\msun$, common at high redshifts, the extent of the shock
radius is likely to decrease in rare higher-mass halos, with
$\gtrsim10^7\msun$, due to their deeper potential wells and the effect
of infalling gas, possibly trapping the I-front within the host halo,
resulting in a compact \ion{H}{2} region \citep{Kitayama2004,Yoshida2007a}.

At the end of the life of the Pop~III star after $\sim$2.7 Myr, when it leaves a massive BH remnant behind, the surrounding gas has been evacuated from the minihalo, thus preventing the relic BH from accreting any cold gas for an extended period of time. After the radiation from the central star shuts down, the relic \ion{H}{2} region starts cooling partly due to adiabatic expansion cooling, and partly due to atomic hydrogen line cooling, facilitated by collisional excitations from the enhanced abundance of free electrons. The elevated electron fraction catalyzes the formation of $\rm H_2$ and HD molecules in the relic \ion{H}{2} region, thus providing additional molecular cooling \citep{Ricotti2001, JGB2007}. Roughly 10\,Myr after the radiation from the Pop~III star turns off, the temperature decreases from $10^4$\,K to $10^3$\,K and the $\rm H_2$ abundance increases from $10^{-8}$ to $10^{-4}$. While $\rm H_2$ and HD molecules are susceptible to photo-dissociation from Lyman-Werner (LW) radiation produced in neighboring star-forming halos, the relatively short time required for the re-formation of $\rm H_2$ and HD, compared to the average time between the formation of Pop~III stars, allows their abundances to remain high.

\begin{figure}[ht]
\epsscale{1.0}
\plotone{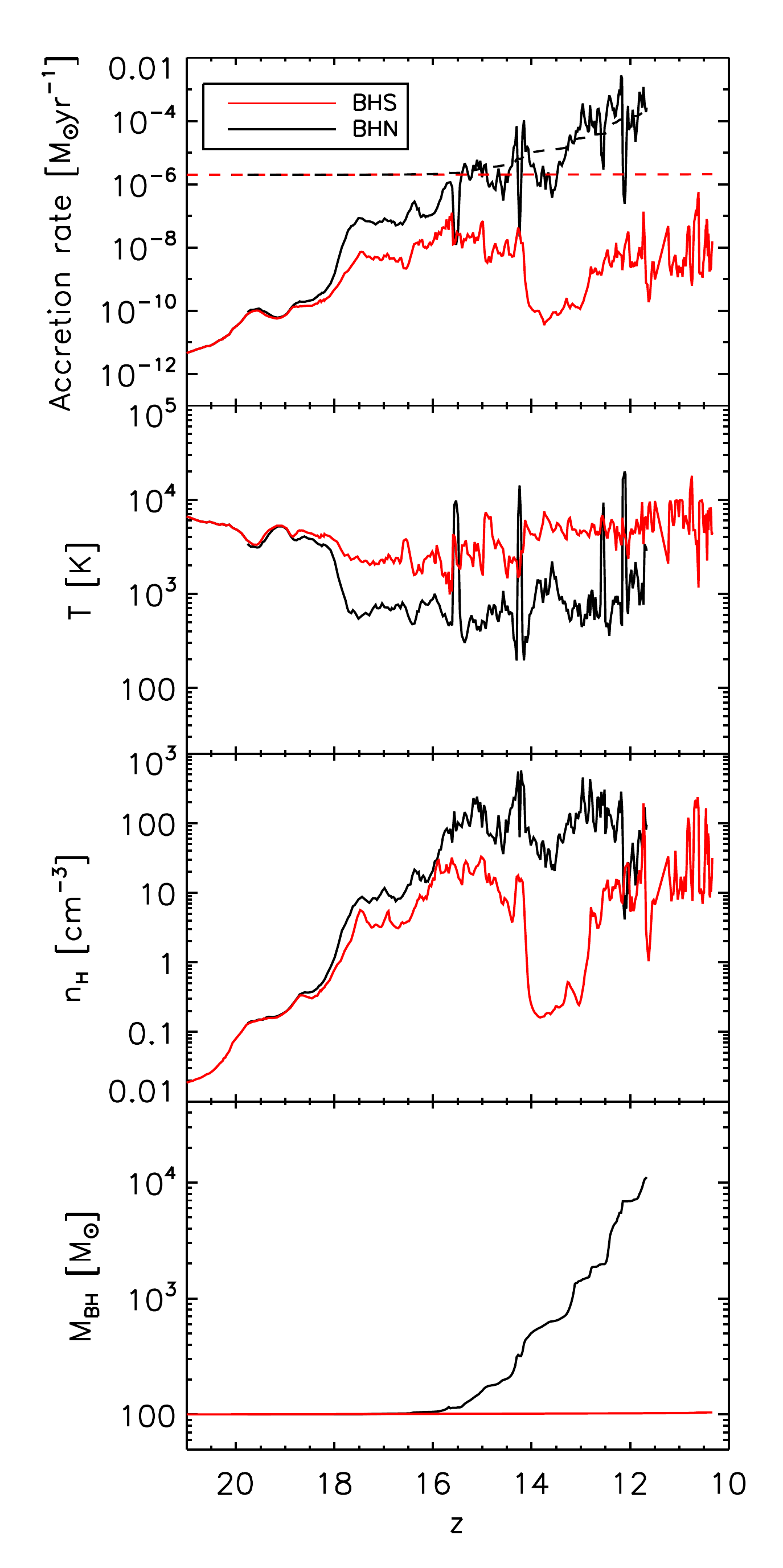}
\caption{BH growth with and without feedback. Shown is the redshift evolution of the BH accretion rate, the temperature and density of the gas in the immediate vicinity of the BH, as well as the resulting BH masses for simulations BHN (\emph{black}) and BHS (\emph{red}). In the top panel, we also indicate the
corresponding Eddington-limited accretion rates for the two cases (\emph{dashed lines}).}
\label{accrete}
\end{figure}

\subsection{Star Formation}
\label{s32}

Figure~\ref{sf} shows the distance between newly formed
Pop~III stars and the central BH in the range of redshifts
$z=10-20$ from three simulations. A total of $\sim50$ Pop~III stars have been formed in the BHN and BHS cases, accompanied
by individual \ion{H}{2} regions, according to our criterion that
stellar radiative feedback is taken into account only for those stars
formed within a 10 kpc radius from the center of the emerging
protogalaxy. We find that for over 250\,Myr after the seed BH formed
there has been no further star formation within the host halo in simulation BHS. This is 
because it takes time for the gas expelled by the BH progenitor star to be
reincorporated into the halo, and the modest feedback from the BH
prevents the gas from cooling. In simulation BHN, on the other hand, stars inside the host halo 
continue to form already much earlier, demonstrating that X-ray heating from the accreting BH 
in simulation BHS implies a strong local negative feedback.

Note the intense burst of star
formation taking place within the virial radius of the protogalactic
halo at $z=11.5-10.5$ in the BHS. This starburst is fueled by a massive infall
of gas. This gas is able to cool not only via molecular hydrogen cooling but also via 
atomic hydrogen line cooling. A fraction of the gas mass
moving toward the central BH is thus consumed by star formation rather than being accreted onto the BH.

In Figure~\ref{fig:sfr}, we show the evolution of the stellar mass
density and the star formation rate density (SFRD). The comoving
stellar mass density is calculated using the stellar masses that
have been formed within the 10 kpc radius from the center of the
emerging protogalaxy at a given redshift. For the SFRD, we take the
time derivative of the stellar mass density at the redshift when the
Pop~III stars formed. For comparison, we overplot SFRDs using analytic
fitting formulae \citep{Hernquist2003}, for higher-mass halos
where atomic hydrogen cooling is dominant, and for minihalos, assuming
that one $100 \msun$ Pop~III star forms per system via $\rm H_2$
cooling \citep{Yoshida2003}.

The estimated star formation rate densities from our simulations lie
between the two analytical fits. Evidently, at $z=20$ the SFRD in the
BHB run is higher than in the BHN and BHS cases, by a factor of about
$3$. This is a consequence of the positive feedback, where gas
collapse into distant minihalos is facilitated
via $\rm H_2$ cooling promoted by the strong X-ray emission from the
HMXB. As density fluctuations grow and star formation
is enhanced, the local negative radiative feedback from Pop~III stars
begins to dominate, mitigating the positive HMXB feedback
effect. We have not considered cooling due to metals and dust, both
produced in SN explosions. At low densities, $n< 500$ $\rm cm^{-3}$,
and at low metallicities, $Z<10^{-3}\zsun$, however, $\rm H_2$ cooling
is expected to dominate over metal-line cooling \citep{Jappsen2007,Jappsen2009a, Jappsen2009b}.

\begin{figure*}[hbt]
\plottwo{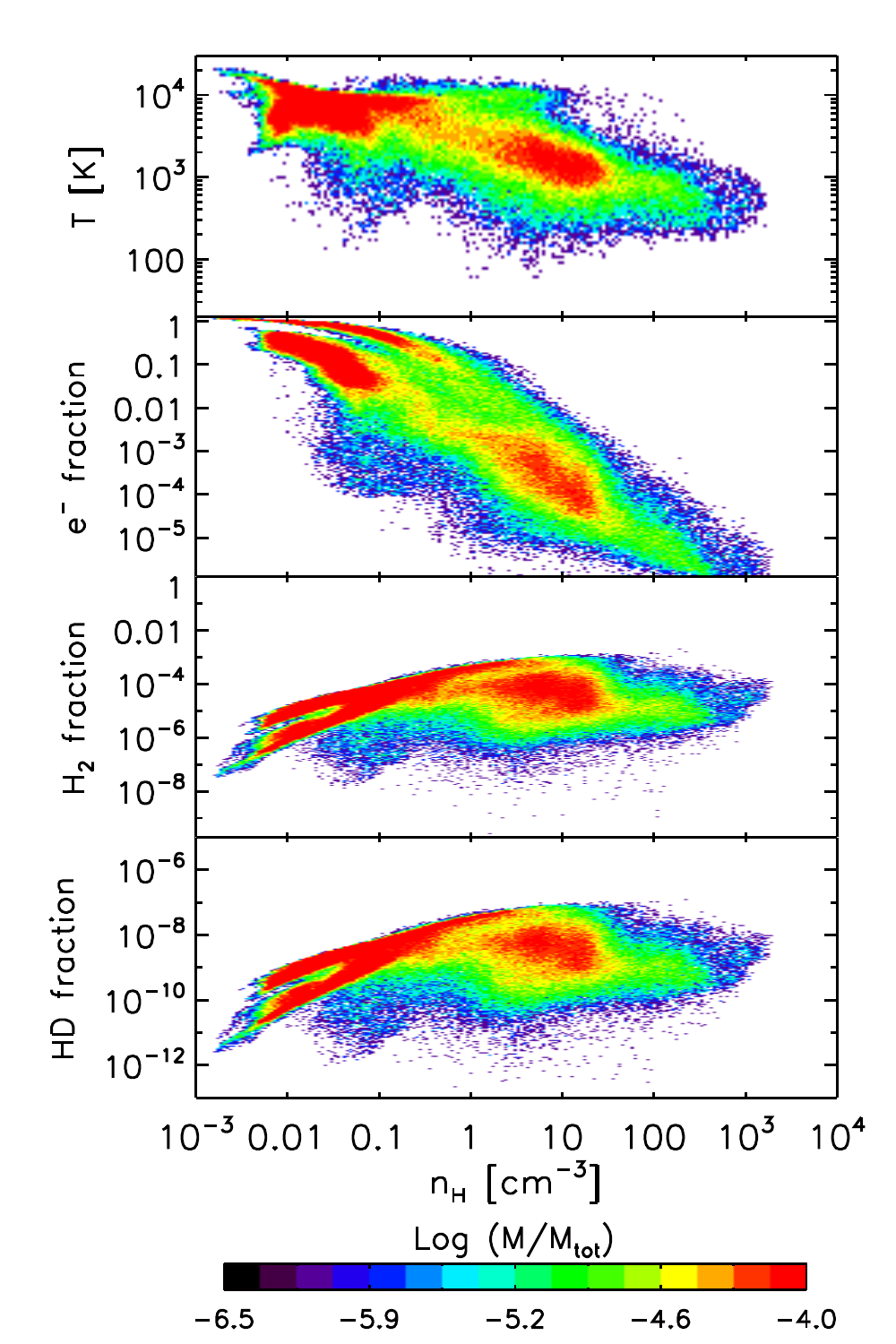}{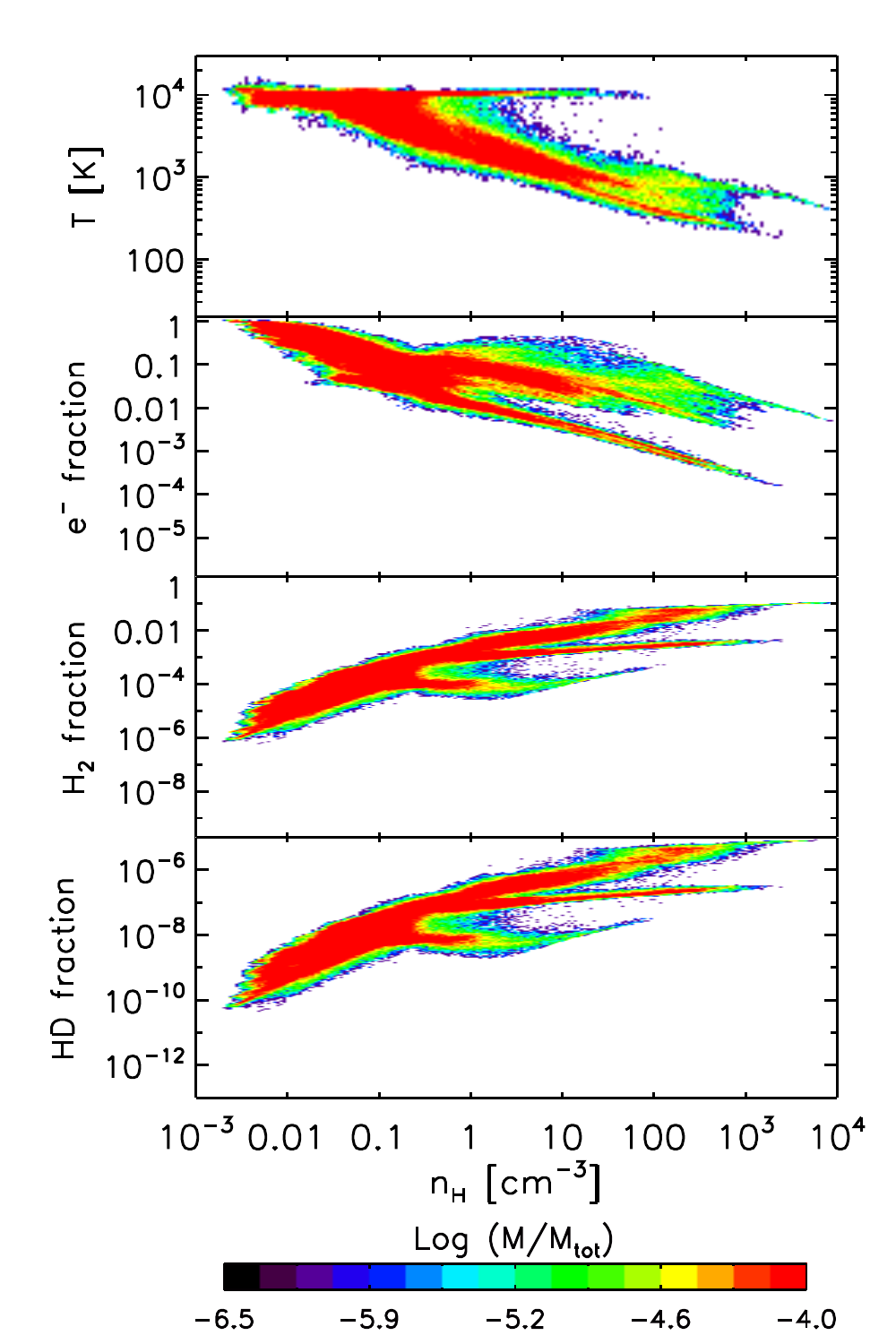}
\caption{Gas properties within the virial radius of the first galaxy halo. We compare the
situation in simulations BHN at $z=11.6$ (\emph{left-hand}) and BHS at $z=10.4$ (\emph{right-hand}), corresponding, respectively, to the end of each of the two simulations. From top to bottom, each panel shows the gas temperature, electron fraction, $\rm H_2$ abundance, and HD abundance, respectively. The feedback from the isolated BH significantly modifies the thermodynamic properties of the central gas. The color coding in these diagrams indicates the fraction of mass in a given phase.}
\label{BHS}
\end{figure*}

\begin{figure}[t]
\epsscale{1.1}
\plotone{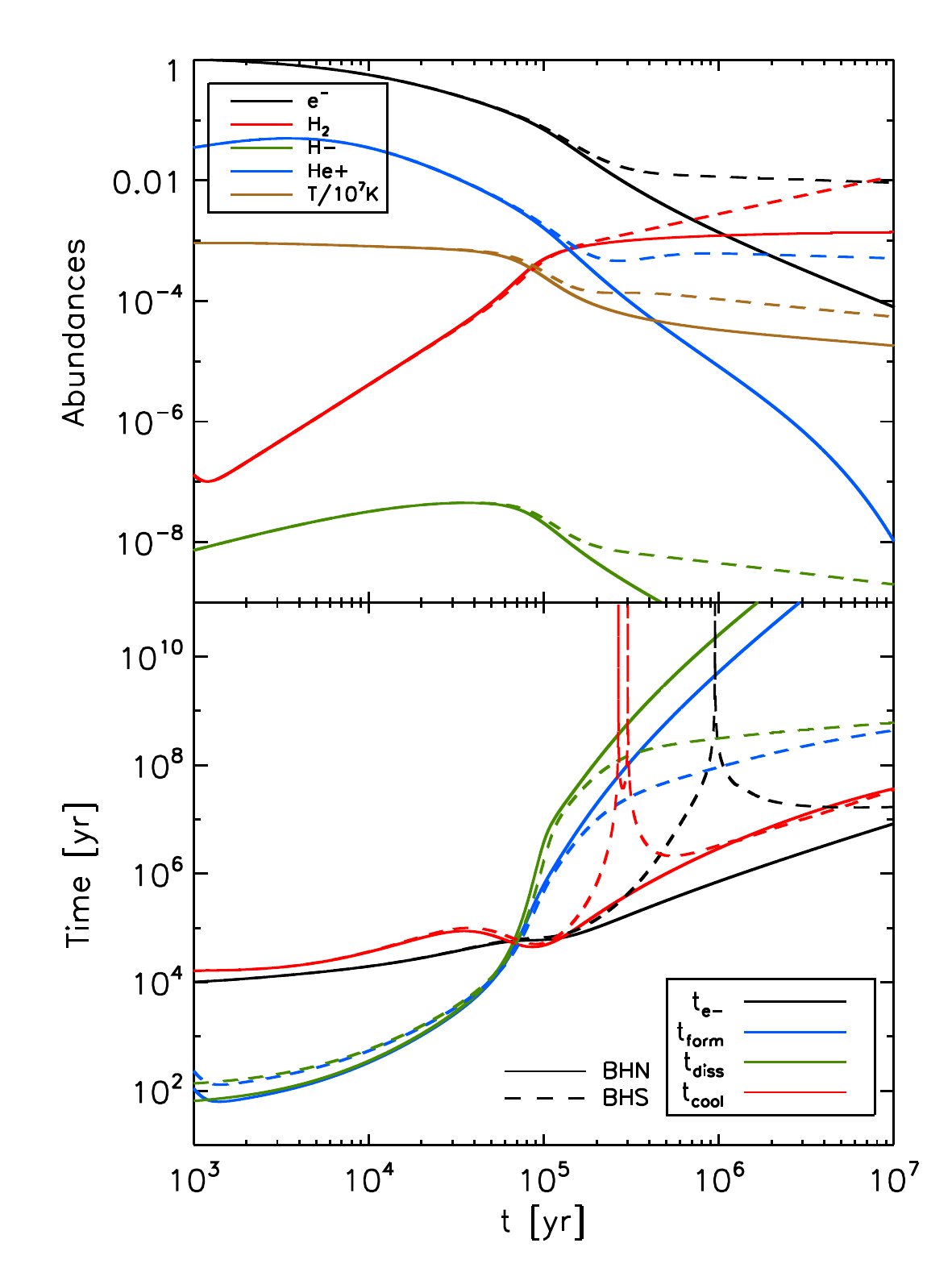}
\caption{Chemical evolution of the central gas cloud. We calculate an
idealized one-zone model, tracing the history of a representative
fluid element in the center of the emerging galaxy, starting from
$T>10^4$\,K and subsequently cooling down. We employ the same
chemistry module as used in the three-dimensional simulations. \emph{Top
panel}: evolution of the particle abundances. \emph{Bottom panel}:
Characteristic timescales. In the absence of feedback from a BH (simulation BHN; solid curves), the $\rm
H_2$ abundance converges to the asymptotic value of $f_{\rm
H_2}\sim10^{-3}$, reached when both the timescale for the evolution of
electron abundance and the cooling time become shorter than the $\rm
H_2$ formation and dissociation timescales \citep{Oh2002}. 
With BH feedback present (BHS; dashed curves), 
the electron fraction is enhanced due to the X-ray
photoionization, and the $\rm H_2$ fraction is enhanced, reaching up
to $\gtrsim10^{-2}$. We note that the additional photo-heating and photo-ionization due to the radiation from a BH increase the cooling time and the timescale for the evolution of the electron fraction, leading to the peaks clearly seen in the bottom panel.}
\label{timescales}
\end{figure}

\begin{figure}[ht]
\epsscale{1.1}
\plotone{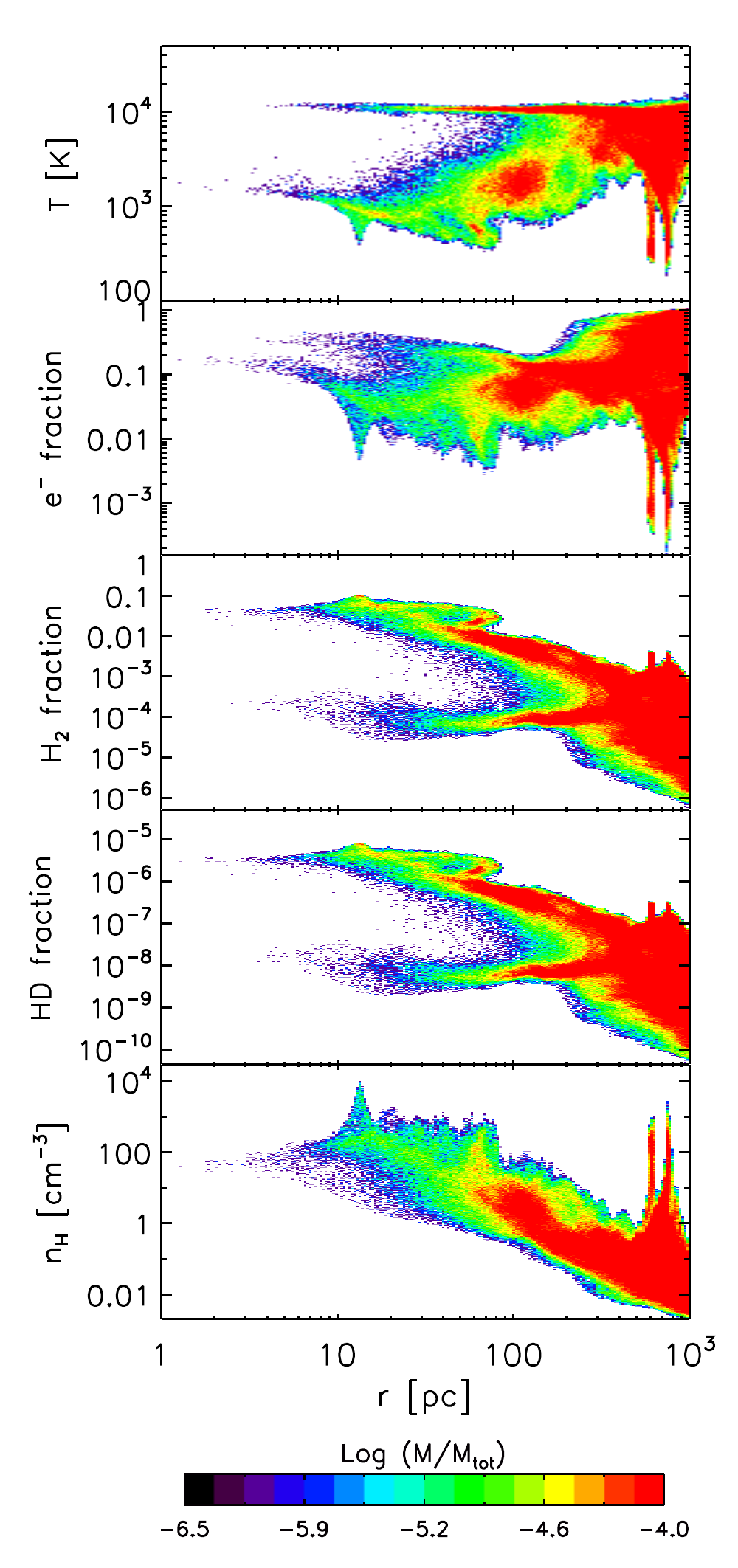}
\caption{Gas properties in the center of the emerging first galaxy. We show temperature, chemical abundances, and number density vs. radial distance from the galactic center for the same simulation BHS at $z=10.4$ as in the right-hand panel of Figure~\ref{BHS}. The radiative feedback from the accreting BH is responsible for the presence of two distinct branches within the central 100\,pc.}
\label{fig:BHS_r}
\end{figure}

\subsection{Mass Growth}
\label{s33}
The characteristic property to distinguish the first galaxies from lower mass minihalos is their
ability to cool through atomic hydrogen line emission, depending on the virial temperature of the hosting halo
with a virial mass $M_{\rm vir}$:
\begin{equation}
\label{virial}
    T_{\rm vir} \sim 10^4 \unit{K} \left(\frac{M_{\rm vir}}{5\times 10^7\msun}\right)^{2/3} \left(\frac{1+z}{10}\right) \mbox{\ .}
\end{equation}

Above $T_{\rm vir} \gtrsim 10^4$ K, the gas within the halos is able
to cool mainly via atomic hydrogen. In Figure~\ref{fig:virial}, we show the evolution
of the virial mass for the most massive halo, which will host the
first galaxy at $z\sim10$, as well as of the total and cold gas masses
for the three simulations presented in this work. The virial mass of
the DM halo is estimated as the mass within a sphere with average
DM density $\rho\sim$ 200 $\rho_0(z)$, where $\rho_0$ is the mean
cosmic density at a given redshift. Cold gas is identified by the
condition that the gas temperature is less than half of the halo
virial temperature, $T<0.5$ $T_{\rm vir}(z)$. We also indicate the
critical mass required for the onset of atomic hydrogen cooling at a
given redshift.

We find that at $z\gtrsim18$, the halo is dominated by hot gas,
exceeding the amount of cold gas by an order of magnitude. As time
passes on, the cold gas mass increases, eventually accounting for
$\gtrsim80\%$ of the total gas mass in simulations BHN and BHS. This
trend can be understood by the vulnerability of the halo gas to
stellar radiative feedback. The corresponding evacuation of gas from
the halo is very strong at high redshifts, $z\gtrsim18$, because the
halo potential wells were not yet deep enough to retain photo-heated
gas. We note that the sharp dips in the amount of cold gas are due to
star formation inside the halo itself (at $z \lesssim 12$), as well as
in neighboring halos that are sufficiently close, within $\lesssim
5$\,kpc. The ionizing feedback from the accreting BH starts operating
at $z<18$, above which the effect is too weak as a result of the
initially very low BH accretion rates.

While the total amount of gas is not sensitive to the BH feedback, the
reduction in cold gas mass by a factor of $\sim 5$ indicates that the
additional heating from this feedback, on top of the stellar feedback,
has a significant impact on the gas in the center of the forming
galaxy. As the halo grows further via smooth accretion and mergers
with minihalos, however, at $z\sim13$, both the total gas mass and mass of 
cold gas are no longer sensitive to the BH radiative
feedback. At $z\sim11.5$, the condition for an atomic cooling halo is
satisfied.

For simulation BHB, the heating from the HMXB is so strong that 
all gas particles have temperatures $T>0.5$ $T_{\rm vir}(z)$,
over the entire range of simulated redshifts $z\gtrsim15$, and the
total gas mass is reduced by an order of magnitude by
photo-evaporation, as is clearly evident in the dot-dashed lines
Figure~\ref{fig:virial}. This result would imply that if an HMXB
existed within a minihalo at high redshifts, it would take
significantly longer for the halo to reassemble the lost gas, and to
eventually evolve into a primordial galaxy.

In Figure~\ref{multi}, we show the evolving conditions of the gas in the vicinity of the accreting BH in simulation BHS, while the first galaxy is assembled. Initially, the BH is located at the center of the \ion{H}{2} region produced by the progenitor star. As filamentary structure develops, gas infall and mergers with minihalos frequently occur along the filaments (see the middle panel of Fig.~\ref{multi}). Cold accretion along the filaments, which are dense enough to allow molecule re-formation, in turn leading to enhanced cooling, efficiently delivers cold gas into the center of the halo. Roughly 300\,Myr after the seed BH formed, the conditions for the onset of atomic hydrogen cooling are met. The virial radius of the halo at this time is $r_{\rm vir}\sim$1 kpc.

\begin{figure}[ht]
\epsscale{1.1}
\plotone{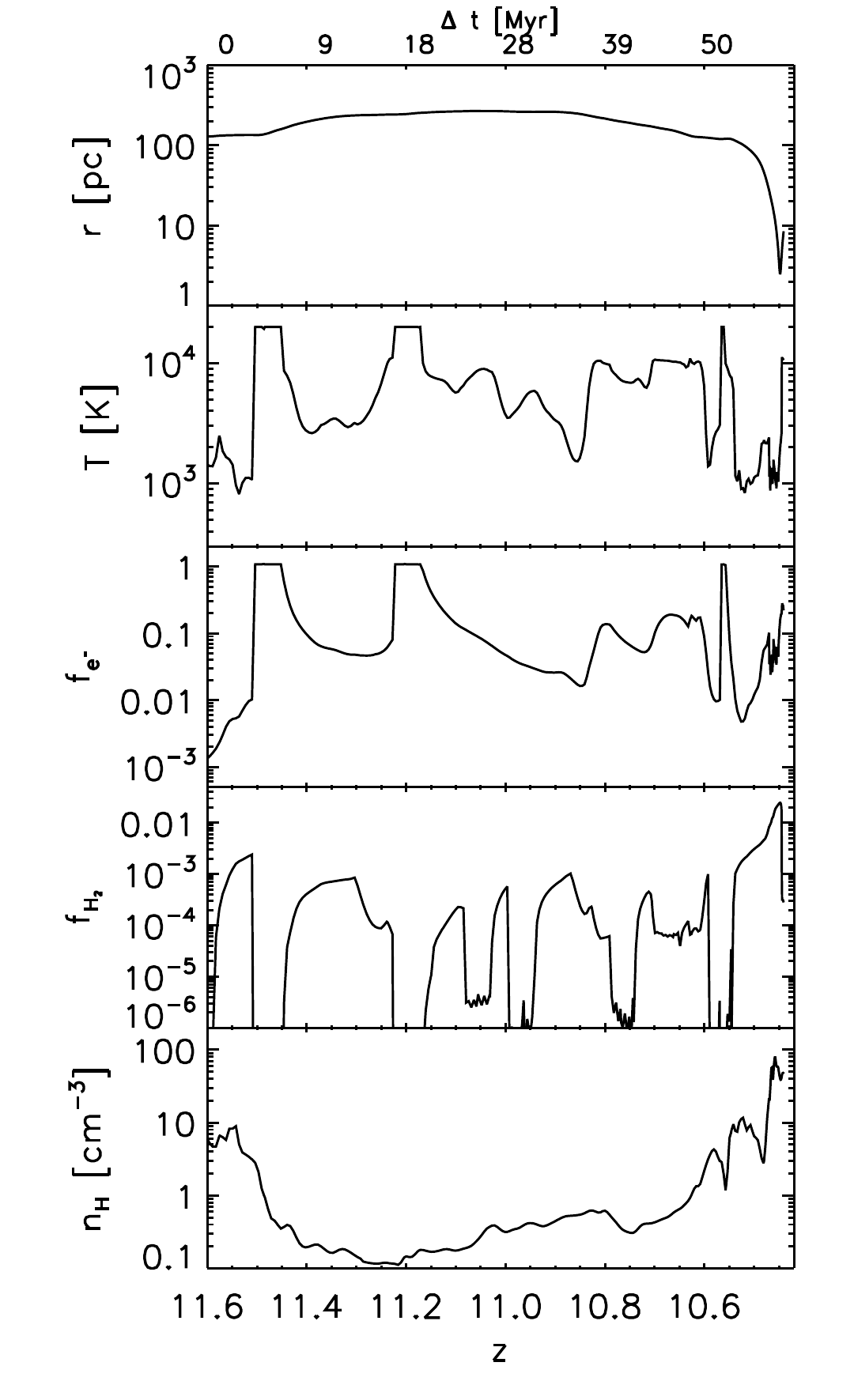}
\caption{History of a representative SPH particle in the vicinity of the central BH. We follow the evolution for 60\,Myr, starting at a distance of 100\,pc away from the BH, all the way to the center of the emerging first galaxy halo. The spikes in the temperature and electron fraction are due to individual star formation events occurring inside or outside of the growing halo. As the gas particle moves to the center of the halo, feedback from the accreting BH begins to operate, promoting $\rm H_2$ and HD molecule formation. Very close to the BH, however, the heating is sufficiently strong to destroy the molecules, generating the two distinct branches for r $\lesssim$ 50 pc seen in Fig.~\ref{fig:BHS_r}.}
\label{partone}
\end{figure}

\subsection{Black Hole Growth}
\label{s34}
Figure~\ref{accrete} shows the evolution of the accretion
rate onto the BH, the temperature and density of the neighboring gas,
as well as the BH mass for simulations BHN and BHS. For the
temperature and density, we average over the 50 closest SPH particles.
The BH remnant, originating in the death of a Pop~III star, initially
resides within a medium of high temperature, $\sim 10^4$\,K, and very
low density, $n_{\rm H}\sim 0.1$\,cm$^{-3}$, preventing the gas from
cooling and hence accreting onto the BH. At this stage, the accretion
rate is only $\dot{M}_{\rm BH}\sim 10^{-11}\msun$\,yr$^{-1}$, and the
corresponding accretion luminosity, assuming a 10$\%$ radiative
efficiency, is $L_{\rm acc}=\epsilon \dot{M}_{\rm BH} c^2 \sim 5.7
\times 10^{33}$ erg\,s$^{-1}$. This value is seven orders of magnitude
below the Eddington luminosity, and BH growth and feedback are 
negligible in the beginning. As the potential well of the host halo
becomes deeper, however, the amount of infalling gas increases with
time, boosting the accretion rate, especially at $z\sim 17-19$, as can
be seen in Figure~\ref{accrete}. Consequently, the accretion luminosity also
becomes large enough to influence the surrounding gas, keeping it at a
temperature that is an order of magnitude higher than in the BHN
comparison simulation. 

The accretion rate in the BHN simulation is already comparable to the
Eddington value, depicted as dashed lines in Figure~\ref{accrete} (top panel), at
$z\sim 15.5$, while it is still an order of magnitude lower in the BHS
case. Occasionally, star formation takes place very close to the BH,
e.g., $\sim 1$\,kpc away at $z\sim 14$ (see Fig.~\ref{sf}). The radiative
feedback from this event acts to compound the heating effect from the
BH accretion, thus rendering the removal of gas out of the shallow
potential well more effective. Such gas evacuation at $z\sim14$ due to
both feedback effects is clearly seen in Figure~\ref{fig:virial}.

The combined stellar and BH radiative feedback results in an accretion
rate that is on average 4 orders of magnitude below the Eddington
value at $z=14-13$. Even 300 Myr after the formation of the BH, its
mass has increased by only 1.5 $\%$ in the BHS simulation, whereas
there is a two orders of magnitude growth in the BHN case. This
indicates that the feedback from a stellar-mass BH is sufficiently
strong to prevent significant growth, suggesting a very important
constraint on SMBH formation scenarios. We infer that the radiative
feedback from an accreting BH might be partly responsible for the low
density of quasars at redshifts $z\sim6$, by suppressing early BH
growth.

\section{Protogalactic Gas Properties}
\label{PGP}

In this section, we discuss the properties of the gas as it falls into the center of the emerging galaxy at $z\sim 10$. We compare the situation for our different assumptions on the feedback acting during the preceding assembly, including a discussion of the complementary PISN case. This allows us to constrain the initial conditions for the subsequent starburst inside the first galaxies, a crucial input for predicting their observational signature.

\begin{figure*}[hbt]
\plottwo{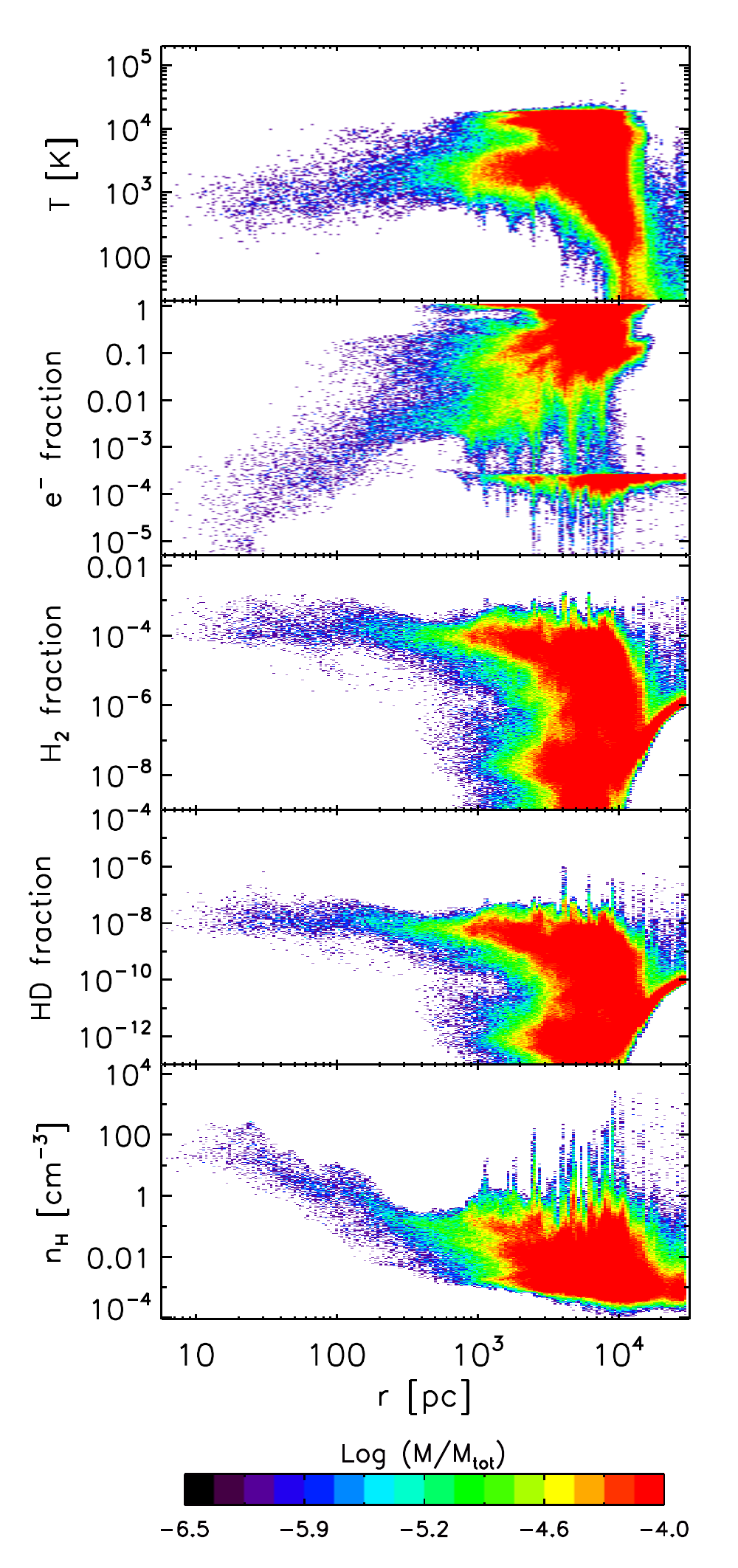}{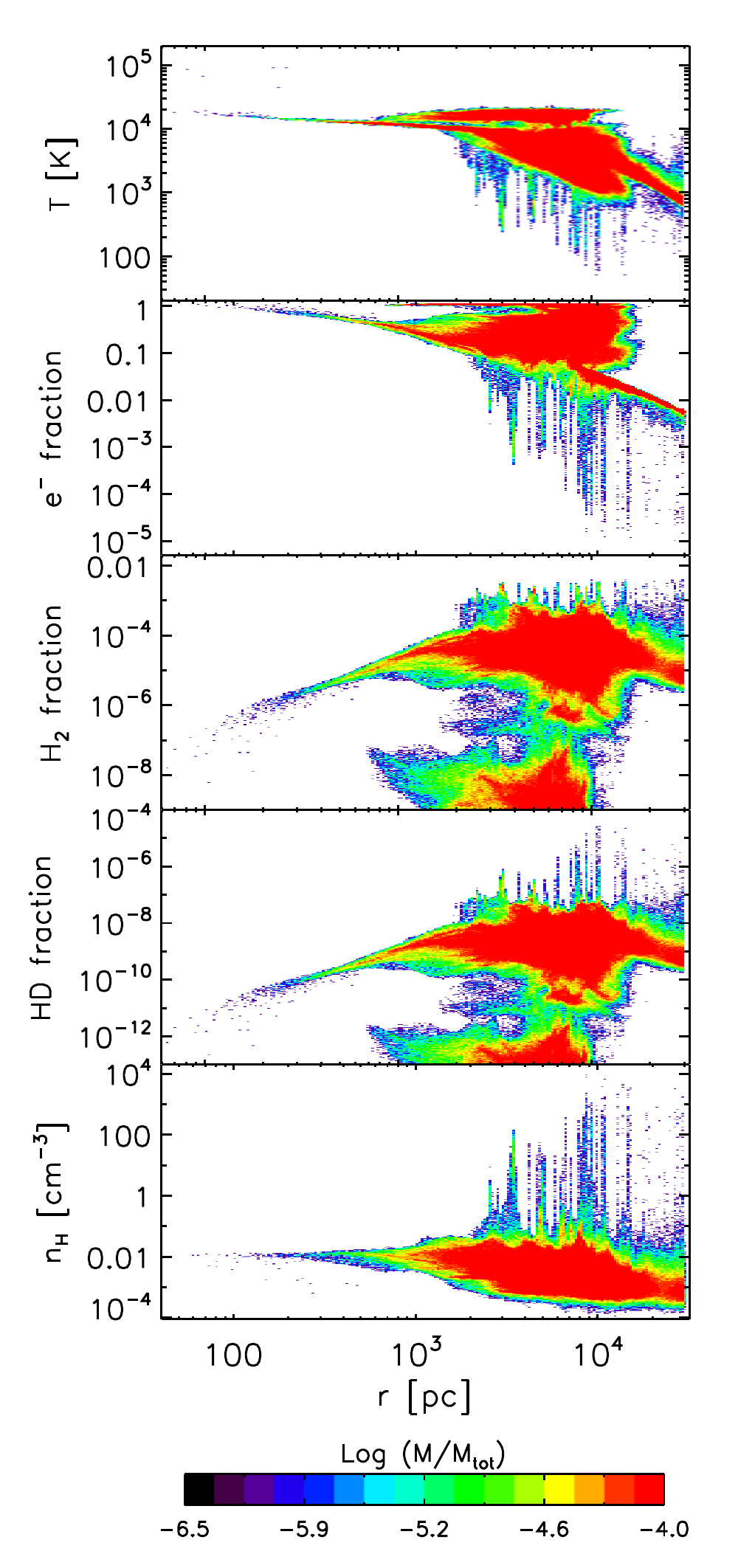}
\caption{Impact of HMXB feedback. We compare the protogalactic gas properties in simulations BHN (\emph{left}) and BHB (\emph{right}) at $z=15$, similar to Fig.~\ref{fig:BHS_r}, but now extending to 30\,kpc to illustrate the HMXB feedback effect not only on the gas within the host halo, but also on the extended IGM.}
\label{BHB}
\end{figure*}

\subsection{Isolated Black Hole Case}

In Figure~\ref{BHS}, we compare the temperature, electron fraction, $\rm H_2$
 fraction, and HD fraction of the primordial gas within the virial
 radius of the first galaxy halo as a function of number density for
 simulations BHN and BHS. We see that at low number densities, $n_{\rm
 H}=10^{-3}-1$ $\rm cm^{-3}$, corresponding to the outskirts of the
 host halo, for both the BHN and BHS simulation the gas temperature is
 high, $T\sim 10^3-10^4$\,K, and the electron fraction is enhanced,
 $f_{e^{-}}=10^{-2}$, due to the accumulated relic \ion{H}{2} regions
 produced by previous star formation inside and outside of the
 halo, and due to heating in accretion shocks. 

 In the BHS run, additional heating from the BH allows the gas
 within the halo to maintain a relatively high temperature over the
 entire range of densities. There is a substantial difference seen in the electron
 fraction: while varying over the large range of $f_{e^{-}}=10^{-6}-1$
 in the BHN simulation, in the BHS case most gas  
 within the virial radius has electron fractions above
 $\gtrsim 10^{-2}$, and there is no gas present with fractions  $ <10^{-4}$. 
 In response to the substantially increased electron
 fraction, the formation of $\rm H_2$ and $\rm HD$ is catalyzed,
 leading to molecule abundances that are higher by two orders of
 magnitude compared to the BHN run.

In the absence of feedback from a BH in simulation BHN, the molecular
hydrogen fraction converges to the well-known asymptotic value of
$f_{\rm H_2}\sim 10^{-3}$. This behavior reflects a freeze-out process \citep{Oh2002}, 
where the molecular hydrogen abundance no longer
evolves once the recombination and cooling timescales become less than
the $\rm H_2$ formation and dissociation timescales, $t_{\rm rec},
t_{\rm cool} \ll t_{\rm form}, t_{\rm diss}$. Here, the
recombination time is $t_{\rm rec}\sim 1/(\alpha_B n f_{e^-})$, and
the $\rm H_2$ formation and dissociation timescales are $t_{\rm
form}\sim n_{\rm H_2}/\dot{n}_{\rm H_2,form}$ and $t_{\rm diss}\sim
n_{\rm H_2}/\dot{n}_{\rm H_2,diss}$, respectively. 
\par
To illustrate the
basic difference between the BHN and BHS simulations, we calculate a
simple one-zone model, using the same chemistry module as in the
simulations, but focusing on a single, representative fluid element as
it falls into the center of the emerging galaxy. In the model, the
radiation from the BH is imposed by a $1/r^2$ radiation field, adopting a typical accretion rate of $10^{-8}\msun \rm yr^{-1}$, and we
assume that the fluid element has a fixed density of $n=10$ $\rm
cm^{-3}$, placed $r=100$\,pc away from the BH at $z=11.6$. Our fluid
element is initially hot, $T=10^4$ K, and highly ionized, typical for
a location inside a relic H~II region. The results of the one-zone model
are shown in the top panel of Figure~\ref{timescales}. To guide the interpretation of these 
results, we also show, in the bottom panel of Figure~\ref{timescales}, 
the cooling time and the timescales for formation and dissociation of molecular hydrogen, 
and for the evolution of the electron fraction, $t_{\rm e}\equiv
n_{e}/\dot{n}_{\rm e}$.  
\par
In the case without BH feedback (solid lines), the $\rm H_2$ formation
and dissociation timescales are significantly longer than both the
timescales for the evolution of the electron fraction and the cooling
rate once the temperature has dropped below a few $10^3$\,K,
establishing the freeze-out value of $f_{\rm H_2}\sim 10^{-3}$. In the
presence of BH feedback (dashed lines), on the other hand, the
additional ionization from the BH X-ray emission significantly
increases the fraction of free electrons at low temperatures. The
enhanced electron fraction promotes the formation of $\rm H^{-}$, and
this, at low temperatures, in turn implies an increase in the $\rm
H_2$ fraction. We note that similarly high $\rm H_2$ fractions have been found in the
related but different context of the non-equilibrium chemistry and cooling of
primordial gas behind structure formation shocks under 
external irradiation \citep{Shapiro1987}.

The overall distribution of the HD abundance is similar to that of
$\rm H_2$. In the BHS run, the HD fraction increases to $f_{\rm
HD}\sim 10^{-6}$ which exceeds the critical level needed for efficient
cooling to the cosmic microwave background (CMB) temperature in local
thermodynamic equilibrium, $f_{\rm HD, crit}\sim10^{-8}$, by two
orders of magnitude \citep{Johnson2006}, as can be seen in
Figure~\ref{BHS}. Nevertheless, HD cooling does not succeed in tying the
temperature to the CMB because of the continuous heating from the BH.

We plot the properties of the gas in simulation BHS in Figure~\ref{fig:BHS_r}, but
this time as a function of distance from the central BH. The most
noticeable feature here is the presence of the two distinct branches
towards the center r $ < 100$ pc which is explained by the BH
feedback. To better understand the BH feedback effect, in Figure~\ref{partone} we
follow the evolution of one representative SPH particle for $\sim
60$\,Myr. Initially, the tracked particle is located at 100\,pc away
from the central BH, where BH feedback is negligible. The sudden jumps
in both temperature and electron fraction indicate a number of star
formation events inside and outside of the halo. Soon after such
events, the gas begins to cool back to a few $\sim10^3$\,K, as is
typical for relic \ion{H}{2} regions. 
\par
As the gas particle moves within
100\,pc from the BH, however, the continuous heating from the
accreting BH starts to operate, resulting in temperatures of
$\sim10^3$\,K and an electron fraction of $f_{e^-}\sim0.1$. As a
consequence, $\rm H_2$ molecule formation is catalyzed, leading to
fractions of $f_{\rm H_2}\gtrsim 10^{-2}$. Finally, when the gas
particle is located too close to the BH, $r \lesssim$ 10\,pc, the
strong X-ray flux heats up the gas again to $\sim10^4$\,K, leading to
the dissociation of molecules and producing the two distinct branches
within the central 100\,pc, clearly seen in Figure~\ref{fig:BHS_r}. We point out
that the dense clumps around $700-800$\,pc from the BH in Figure~\ref{fig:BHS_r} correspond to the middle branch in the $\rm H_2$ and HD
molecule fractions, evident in the right panel of Figure~\ref{BHS}.

\begin{figure}[t]
\epsscale{1.15}
\plotone{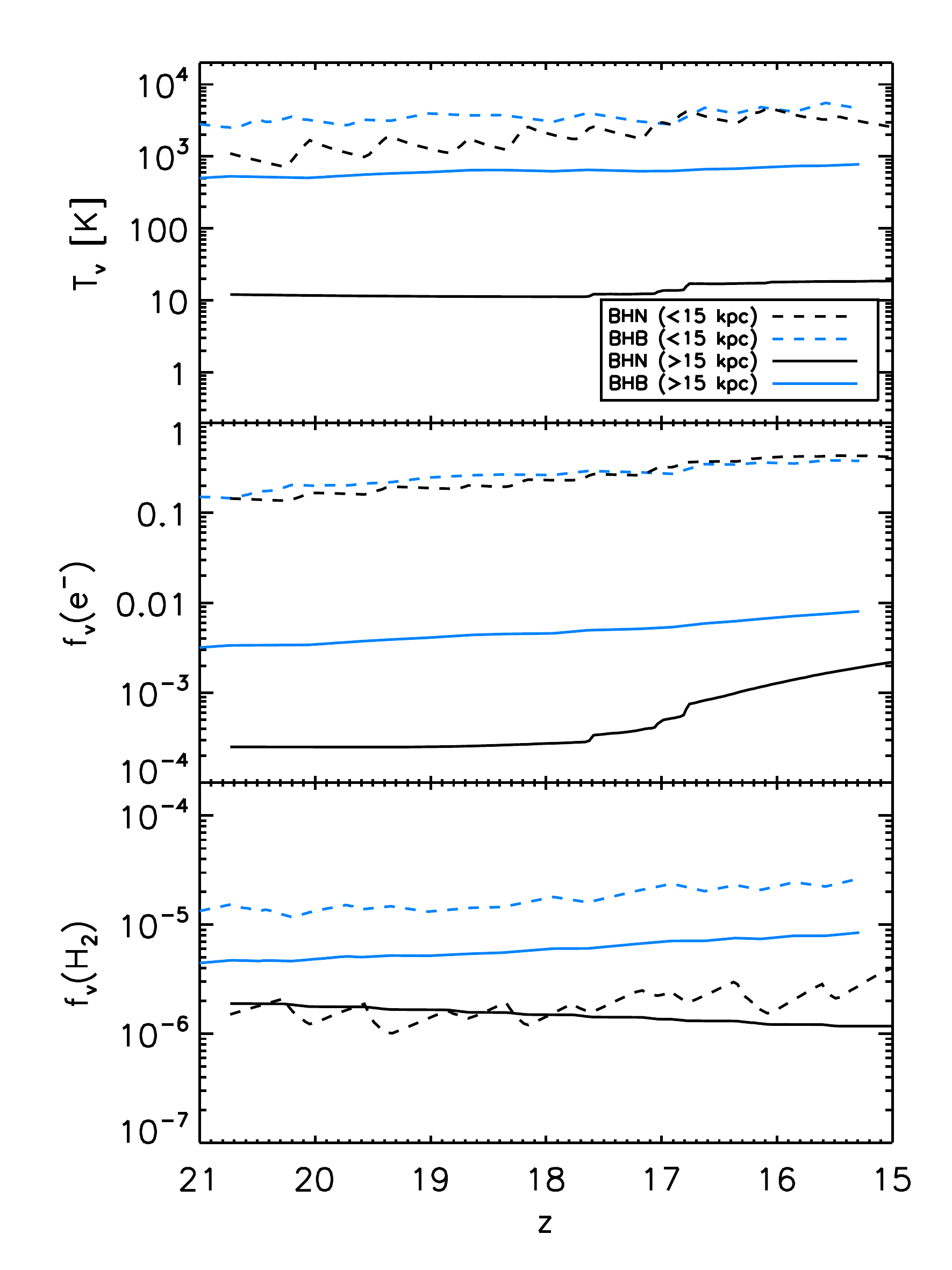}
\caption{Impact of HMXB feedback on IGM. We show the volume-averaged temperature, electron fraction, and $\rm H_2$ fraction vs. redshift in simulations BHN and BHB. We show these quantities separately for radial distances smaller and larger than 15\,kpc, beyond which the feedback from Pop~III stars is not included. Hence, the feedback effect from the HMXB on the distant IGM becomes clearly apparent.}
\label{BHB_ion}
\end{figure}

\subsection{Black Hole Binary Case}
In simulation BHB, where the continuous feedback from an HMXB is included, a dramatic change is seen in the gas properties surrounding the central HMXB. Figure~\ref{BHB} shows the comparison between simulations BHN and BHB at $z=15$, plotted in the same fashion as in Figure~\ref{fig:BHS_r}, but now extending to 30\,kpc from the HMXB to illustrate the effect of feedback not only on the gas within the host halo, but also on the distant IGM. We carefully select the snapshots to guarantee that we are seeing the feedback effect from the HMXB, without confusion from Pop~III stellar radiation. In the BHB run, the gas temperature within the central 1\,kpc reaches $\sim 10^4$\,K, or even higher. The $\rm H_2$ and HD molecule abundances are consequently suppressed via effective collisional dissociation, leading to values of $f_{\rm H_2}\sim10^{-6}$ and $f_{\rm HD}\sim10^{-11}$, respectively. The mean density of the gas in the central 100\,pc is $n_{\rm H}\sim 0.01$\,cm$^{-3}$ which is 4 orders of magnitude smaller than that in the BHN run, showing the strong photoevaporation due to HMXB radiative feedback. 

In both simulations part of the IGM outside a 1\,kpc radius is heated and ionized by an I-front which had been generated 6.5\,Myr ago by Pop~III stars formed $\sim 5$\,kpc away from the center of the halo hosting the HMXB. The HMXB feedback effect on the IGM is most clearly evident in gas which has never experienced any stellar radiation feedback, as the latter could masquerade as the former. The distant IGM in the BHN simulation is in a cold and neutral state, while in the BHB run, the IGM unaffected by any stellar radiative feedback is in a warm and partially ionized phase because of the pervasive effect of the strong HMXB emission. These are favorable conditions for $\rm H_2$ and HD molecule formation, promoting their abundance by an order of magnitude over the BHN case.

To more clearly distinguish the HMXB feedback on the first galaxy and on the distant IGM, we compare in Figure~\ref{BHB_ion} the volume-averaged temperature, electron fraction, and $\rm H_2$ fraction for simulations BHN and BHB inside and outside of a 15\,kpc radius, beyond which our algorithm does not trigger the build-up of \ion{H}{2} regions around newly-formed Pop~III stars. Hence, any variations in gas properties in the far-zone are caused by the HMXB feedback alone. Here, we adopt a 15\,kpc cutoff radius rather than 10\,kpc, the region inside which stars are endowed with an \ion{H}{2} region (see Sec.~\ref{s24}), because an ionized region generated by a Pop~III star formed near $r\sim$ 10\,kpc from the center of the emerging galaxy will extend out to $\sim 15$\,kpc. The corresponding X-ray heating increases the volume-averaged temperature within 15\,kpc by only a factor of 3, while the temperature in the distant IGM is enhanced by two orders of magnitude. The difference in the near zone is relatively small because the gas within the 15\,kpc radius is affected by both the feedback from Pop~III stars, and from the HMXB, indicating that the overall gas properties are dominated by the local stellar feedback. In the absence of any local stellar feedback, as is the case in the region beyond the 15\,kpc radius, on the other hand, the IGM is substantially pre-heated, and partially ionized when HMXB feedback is present. These conditions facilitate molecule formation, enhancing abundances by an order of magnitude. Our results are in good agreement with the study of \citet{Kuhlen2005}, given that the X-ray flux in our simulation BHB is comparable to that for their power-law case. 

We assume that the HMXB source has been turned on for $\sim150$\,Myr, which is much longer than any realistic lifetimes. For example, the main-sequence lifetime of a donor star of $30\msun$, close to the predicted typical Pop~III mass \citep{Clark2011b, Hosokawa2011, Stacy2012}, would be $\sim 20$\,Myr. Thus, our work can be considered to represent the case where at least one HMXB exists continuously over 150 Myr, somewhere in the Lagrangian volume of the emerging protogalaxy. A preheated IGM would provide us with a promising observational window into the early universe. Under these conditions, the IGM would be sufficiently hot to be decoupled from the CMB temperature, $T_{\rm CMB}=2.73$\,K\,$ (1+z)$, but not hot enough to be ionized, leaving the IGM substantially neutral. Such a preheated, neutral IGM is likely detectable in the emission of redshifted 21\,cm radiation, which is produced by the spin-flip transition between the singlet and triplet hyperfine- structure levels of neutral hydrogen. In principle, future 21\,cm observations could provide constraints on the character of early feedback processes, and in particular on the importance of a contribution from Pop~III miniquasars \citep[see e.g.][]{Furlanetto2006}.

\subsection{Comparison to SN Feedback}
In the previous section we discussed the situation established under radiative feedback from an accreting isolated BH, and an HMXB. Effectively, we have made the assumption that all Pop~III stars end their lives as a BH, without any SN explosion preceding their deaths. If a Pop~III star, however, has a mass in the range of $140-260\msun$, it would die as a PISN, exerting a strong mechanical feedback on the surrounding primordial gas, enriching it with the metals produced before and during the explosion. We are able to directly compare our results to those of \citet{Greif2010}, where PISN enrichment of the IGM, stellar radiative feedback, chemical mixing, and metal-line cooling were included, because all simulations start from the same initial conditions. The underlying DM structure was thus identical, providing us with an ideal laboratory to study the impact of variations in the baryonic physics on the assembly process of the first galaxies. 

We find that the gas properties within the virial radius of the host halo in our BHN simulation at $z\sim10$ are very similar to those in the \citet{Greif2010} PISN simulation, except that the central gas in the latter simulation is enriched to $Z\sim10^{-3}\zsun$. It is, however, still unclear whether such enrichment level is sufficient to enable the transition from Pop~III to Pop~II star formation, because the critical metallicity threshold for this transition is still a matter of debate. The additional heating from the central BH in our BHS run, on the other hand, significantly enhances molecule formation, possibly by up to two orders of magnitude, and the gas might be able to further fragment. {\it Which feedback is dominant in shaping second-generation star formation in the first galaxies?} The answer will strongly depend on the IMF of the first generation of stars, and on the detailed physics of the transition from the predicted top-heavy IMF for Pop~III to the more normal IMF in already metal-enriched gas.

\section{Summary and Conclusions}
\label{SC}

We have studied how the assembly of a primordial galaxy is affected by
the radiative feedback from an accreting, isolated stellar-mass black hole and
a high-mass X-ray binary (HMXB), which are two possible end products of Pop~III star
formation. To accomplish this, we have carried out three cosmological
simulations which self-consistently account for the radiation from
individual Pop~III stars, and from a central black hole X-ray source. In
particular, we have focused on the early growth of a Pop~III black hole 
remnant at high redshifts, and on the role of black hole feedback in shaping
subsequent star formation in the emerging first galaxy at
$z\sim10$. Our main results and implications can be summarized as
follows.
\\

1. We have shown that locally, as opposed to any global effect on the large-scale IGM, the feedback from an isolated, accreting black hole is very efficient, leading to a strong suppression of the early growth of the seed black hole. Without such feedback, the growth rate quickly reaches near-Eddington values. Our results imply the following important consequence. A widely adopted assumption for the formation of early supermassive black holes (SMBHs), whose presence is inferred from QSO observations at $z\sim6$, is that their growth begins with a Pop~III seed black hole. If this assumption were correct, quenching of black hole accretion would be required to avoid an overabundance of SMBHs at low redshifts \citep[e.g.,][]{Bromm2003, Trenti2007}. Without such quenching, one would also run into inconsistencies with the low number density of quasars, $(6\pm2)\times10^{-10} \rm Mpc^{-3}$ at $z\sim 6$, assuming $H_0=65 \unit{km} \unit{s}^{-1} \unit{Mpc}^{-1}, \Omega_{m}=0.35$, and $\Omega_{\Lambda}=0.65$ \citep{Fan2004}. We suggest that the radiative feedback from accreting black holes plays a key role in suppressing early black hole growth, thus constraining models for SMBH formation. 

However, in this study we have tracked only an isolated seed black hole, born in a halo of
$5\times10^5\msun$ that virialized at $z=28$. Alternatively, a black hole seed
could form significantly earlier, in a halo corresponding to rare,
very high-$\sigma$ peaks, thus providing a head start to the build-up
process. A different growth-boosting mechanism has been suggested by \citet{Volonteri2005}, 
arguing that Pop~III remnant black holes, hosted in
halos with a virial temperature of $T_{\rm vir}\gtrsim10^4$\,K at
$z\sim 25$, corresponding to peaks of $4\sigma$ or above, may be able
to experience stable supercritical accretion owing to effective atomic
hydrogen line cooling \citep[see also][]{Li2011}. It is still challenging to
numerically test this scenario because of the prohibitive
computational expense to capture such rare massive halos and follow
their evolution.

2. Despite the extremely small accretion rates onto the isolated black hole, the corresponding radiative feedback on the thermal and chemical evolution of the gas in the immediate surroundings, at scales $\lesssim 100$\,pc, is significant. The black hole feedback establishes two distinct thermodynamic branches in the central region of the emerging galaxy: A hot phase with $T\sim 10^4$\,K and high ionization fraction, as well as a warm one with $T\sim 10^3$\,K and partially ionized conditions. In the latter phase, molecular hydrogen formation is favored, leading to an elevated fraction of $f_{\rm H_2}\gtrsim 10^{-2}$, two orders of magnitude higher than in the absence of the radiation from a black hole. This enhanced $\rm H_2$ fraction could promote second generation star formation within the emerging first galaxies, possibly resulting in a burst-like mode. However, we do not include the effect of Lyman-Werner molecule-dissociating photons emitted by the soft (MCD) component of the accreting black hole. Our $\rm H_2$ and HD molecule abundances are therefore upper limits. In future work, we will explore the impact of Lyman-Werner radiation generated by black hole accretion.

What would be the characteristics of the stars formed out of such
metal-free gas with a significantly enhanced electron fraction, the
so-called Pop~III.2 stars \citep{OSheaAIPC}. This issue
is far from being settled. Existing studies suggest that the elevated
$\rm H_2$ and HD fractions will lead to rapid cooling, resulting in
smaller mass stars, of the order of $\sim 10\msun$, compared to the
canonical value of $\sim 100\msun$ for Pop~III.1 stars formed in
quasi-neutral minihalos \citep[e.g.,][]{OShea2005, Johnson2006, Yoshida2007b}. 
However, the suite of simulations carried
out by \citet{Clark2011a}, who considered turbulent fragmentation in
primordial gas, has indicated, somewhat counter-intuitively, that the
second generation, Pop~III.2 stars were likely more massive than the
Pop~III.1 stars, formed in the pristine minihalos that are unaffected
by prior star formation. This can be understood by the inability of
the gas in the Pop~III.2 case to reach the CMB temperature floor via
HD cooling under realistic halo conditions.

3. The feedback from an efficiently radiating HMXB is very strong
   locally, and moderately important globally. Locally, the effect on
   the surrounding primordial gas is to heat it to high temperatures
   of $\gtrsim 10^4$K, and to fully ionize it. The corresponding
   strong photo-evaporative outflow suppresses central gas densities,
   thus preventing any subsequent star formation within the emerging
   galaxy. Our results imply that once a halo of $\sim10^6\msun$
   harbors an HMXB, the ensuing strong radiative feedback will delay
   the condensation of gas in the atomic cooling halo, possibly leading to a decrease in
   the number of first galaxies at a given epoch. While we explore in
   this paper the impact of a continuously radiating HMXB, the
   strength of the effect will depend on the realistic abundance and
   lifetime of such HMXB sources at high redshifts. The highly
   penetrating X-ray photons emitted by the HMXB are able to
   effectively heat and partially ionize the distant IGM as well, and
   as a result catalyze the formation of molecular hydrogen over large
   volumes. We have shown that the global star formation rate in the
   presence of HMXB feedback is, at least at early times, higher than in the runs without it. 
We assume that the broad protostellar mass function seen in recent high-resolution simulations suggesting that 
primordial stars have less extreme masses is likely to
increase the possibility of HMXBs.
However, the mapping between the protostellar fragments, 
as represented by sink particles, to the final Pop~III stars, is still incompletely understood. 
Improved simulations should soon be able to further elucidate this key issue \citep[see][]{Greif2012}.

To sum up, the radiative feedback from accreting black hole sources, either isolated or in a binary system, has a dramatic impact on the surrounding gas within the first galaxy. The alternative, where no black hole feedback is present, but where the gas is processed by a preceding PISN explosion, is less affected in the long run. However, it is likely that the enhanced cooling due to metals and dust will only become effective at densities that are higher than can be resolved in these simulations. One possible consequence of metal and dust cooling might be an enhanced star formation efficiency, potentially enabling the formation of a {\it bona fide} Pop~II, stellar cluster inside the first galaxies \citep{Clark2008, Tsuribe2008, Dopcke2011}. To better understand such Pop~II star formation, we need to follow the subsequent collapse of the dense, metal-enriched gas with much higher numerical resolution in future simulations.
\\

\begin{acknowledgements}
V.~B.\ and M.~M.\ acknowledge support from NSF grants AST-0708795 and AST-1009928 and NASA ATFP grant NNX09AJ33G.  V.~B.\ also acknowledges support from JPL Research Support Agreement 1354840. The simulations were carried out at the Texas Advanced Computing Center (TACC). V.~B.\ thanks the Max-Planck-Institut f\"ur Astrophysik for its hospitality during part of the work on this paper.  R.~S.~K.\ and S.~C.~O.~G.\ acknowledge support by contract research {\em Internationale Spitzenforschung II} of the Baden-W\"urttemberg Stiftung (grant P-LS-SPII/18), from the German {\rm Bundesministerium f\"ur Bildung und Forschung} via the ASTRONET project STAR FORMAT (grant 05A09VHA), from the {\em Deutsche Forschungsgemeinschaft} (DFG) under grants no.\ KL1358/10 and KL1358/11, via the collaborative research project SFB 881 ``The Milky Way Galaxy", and via the priority program SPP 1573 ``Physics of the Interstellar Medium'' (KL 1358/14).
\end{acknowledgements}

\bibliographystyle{apj}
\bibliography{apj-jour,myrefs2}


\end{document}